\documentclass[aps, prd, amsmath, twocolumn, nofootinbib, superscriptaddress, showpacs, eqsecnum]{revtex4}

\usepackage{graphicx}
\usepackage{amsmath,amssymb}
\usepackage{amsfonts}
\usepackage{xspace} 
\usepackage[usenames]{color}
\usepackage{dcolumn}
\usepackage{bm}

\usepackage{ulem}
\newcommand{\Maryland}{\affiliation{Maryland Center for Fundamental
    Physics, Department of Physics, University of Maryland, College
    Park, MD 20742}}

\newcommand{\eq}{\begin{equation}}
\newcommand{\eeq}{\end{equation}}
\newcommand{\be}{\begin{equation}}
\newcommand{\ee}{\end{equation}}
\newcommand{\bea}{\begin{eqnarray}}

\newcommand{\eea}{\end{eqnarray}}
\newcommand{\bes}{\begin{subequations}}
\newcommand{\ees}{\end{subequations}}

\begin{document}

\title{An improved effective-one-body Hamiltonian for spinning black-hole binaries}

\author{Enrico Barausse} \Maryland %
\author{Alessandra Buonanno} \Maryland %

\begin{abstract} 
Building on a recent paper in which we computed the canonical Hamiltonian of a spinning 
test particle in curved spacetime, at linear order in the particle's spin, we work out 
an improved effective-one-body (EOB) Hamiltonian for spinning black-hole binaries. 
As in previous descriptions, we endow the effective particle not only 
with a mass $\mu$, but also with a spin $\mathbf{S}_\ast$. Thus, the effective particle
interacts with the effective Kerr background (having spin 
$\mathbf{S}_{\rm Kerr}$) through a geodesic-type interaction 
and an additional spin-dependent interaction proportional to $\mathbf{S}_\ast$. 
When expanded in post-Newtonian (PN) orders, the EOB Hamiltonian reproduces
the leading order spin-spin coupling and the spin-orbit 
coupling through 2.5PN order, for any mass-ratio. Also, it reproduces
{\it all} spin-orbit couplings in the test-particle limit. Similarly to the 
test-particle limit case, when we restrict the EOB dynamics 
to spins aligned or antialigned with the orbital angular 
momentum, for which circular orbits exist, 
the EOB dynamics has several interesting features, such as the existence of an 
innermost stable circular orbit, a photon circular orbit, and a maximum in the orbital frequency during the
plunge subsequent to the inspiral. These properties are crucial 
for reproducing the dynamics and gravitational-wave emission of 
spinning black-hole binaries, as calculated in numerical relativity simulations.
\end{abstract}

\date{\today \hspace{0.2truecm}}


\maketitle

\section{Introduction}
\label{sec:intro}

Coalescing black-hole binaries are among the most promising sources
for the current and future laser-interferometer gravitational-wave  
detectors, such as the ground-based detectors LIGO and Virgo~\cite{Waldman:2006,Acernese:2008zz} 
and the space-based detector LISA~\cite{schutz_lisa_science}.

The search for gravitational waves from coalescing binaries and the 
extraction of the binary's physical parameters are based on the matched filtering technique, 
which requires accurate knowledge of the waveform of the incoming signal. 
Because black holes in general relativity are uniquely defined by their masses and spins, 
the waveforms for black-hole binaries on a quasi-circular orbits depend on eight parameters, namely
the masses $m_1$ and $m_2$ and the spin vectors $\boldsymbol{S}_1$ and $\boldsymbol{S}_2$. Due to the large 
parameter space, eventually tens of thousands of waveform 
templates may be needed to extract the gravitational-wave 
signal from the noise, an impossible demand for numerical-relativity alone. 
Fortunately, recent work at the interface between analytical and numerical relativity 
has demonstrated the possibility of modeling analytically the dynamics and 
the gravitational-wave emission of coalescing {\it non-spinning} black holes, 
thus providing data analysts with analytical template 
families~\cite{Buonanno2007,Ajith-Babak-Chen-etal:2007b,Damour2009a,Buonanno:2009qa}  
to be used for the searches (see also Ref.~\cite{Yunes:2009ef}, which considers the cases of extreme mass-ratio 
inspirals). The next important step is to extend those studies to {\it spinning precessing} 
black holes.

So far, the analytical modeling of the inspiral, plunge~\footnote{We refer to
plunge as the dynamical phase starting soon after the two-body system passes the last stable orbit. 
During the plunge the motion is driven mostly by the conservative dynamics.}, merger~\footnote{
We refer to merger as the dynamical phase in which the two-body system 
is described by a single black hole.}, 
and ringdown has been obtained within either the effective-one-body (EOB)
formalism~\cite{Buonanno00,Buonanno99,DJS3PN,Buonanno-Cook-Pretorius:2007,
Pan2007,Buonanno2007,Damour2007a,DN2007b,DN2008,Boyle2008a,Damour2009a,Buonanno:2009qa} or 
in Taylor-expanded PN models~\cite{Pan2007}, both calibrated to 
numerical-relativity simulations, or in phenomenological 
approaches~\cite{Ajith-Babak-Chen-etal:2007b,Ajith2009} where the numerical-relativity 
waveforms are fitted to templates which 
resemble the PN expansion, but in which the coefficients predicted 
by PN theory are replaced by many arbitrary coefficients. Considering the success 
of the EOB formalism in understanding the physics of the 
coalescence of non-spinning black holes and modeling their gravitational-wave 
emission with a small number of adjustable parameters, 
in this paper we will use that technique, adapting it to the case of spinning black-hole binaries.

The first EOB Hamiltonian which included 
spin effects was computed in Ref.~\cite{Damour01c}. In Ref.~\cite{Buonanno06}, the authors 
used the non-spinning EOB Hamiltonian augmented with PN spin terms to carry out the 
first exploratory study of the dynamics and gravitational radiation of 
spinning black-hole binaries during inspiral, merger and ringdown. 
More recently, Ref.~\cite{DJS08} extended the model of Ref.~\cite{Damour01c} to include the next-to-leading-order spin-orbit couplings. 
The EOB formalism developed in Refs.~\cite{Damour01c,DJS08} highlights 
several features of the spinning two-body dynamics and was recently 
compared to numerical-relativity simulations of spinning non-precessing black holes in Ref.~\cite{Pan2009}. 
In this paper we build on Refs.~\cite{Damour01c,DJS08} and also on Ref.~\cite{diracbrackets_ours}, 
in which we (in collaboration with Etienne Racine) derived the canonical Hamiltonian for a spinning test-particle
in curved spacetime, at linear order in the particle's spin, and work out 
an improved EOB Hamiltonian for spinning black-hole binaries. In particular, 
our EOB Hamiltonian reproduces
the leading order spin-spin coupling and the spin-orbit 
coupling through 2.5PN order, for any mass-ratio. Also, it
 resums \textit{all} the test-particle limit spin-orbit 
terms. Moreover, when restricted to the case of spins aligned or antialigned with the orbital angular momentum, 
it presents several important features, such as the existence of an 
innermost stable circular orbit, a photon circular orbit, and a maximum in the orbital frequency
during the plunge subsequent to the inspiral. All of these features are crucial 
for reproducing the dynamics and gravitational-wave emission of 
spinning coalescing black holes, as calculated in numerical relativity simulations.

This paper is organized as follows. After presenting our notation (Sec.~\ref{sec:notation}), 
in Sec.~\ref{sec:Hamiltonian_axisymmetric} we build on 
Ref.~\cite{diracbrackets_ours} and derive the Hamiltonian for a spinning test particle 
in axisymmetric stationary spacetimes. In Sec.~\ref{sec:hamiltonian_kerr}, we specialize 
the axisymmetric stationary spacetime to the Kerr spacetime in Boyer-Lindquist coordinates.
In Sec.~\ref{sec:deformed_metric} we work out the EOB Hamiltonian of two spinning precessing 
black holes. In Sec.~\ref{sec:hamiltonian_eob} we restrict the dynamics to spins aligned or antialigned 
with the orbital angular momentum and determine several properties of the circular-orbit 
dynamics. Section~\ref{sec:conclusions} summarizes our main conclusions. 
More details on how the spin-spin sector of the EOB Hamiltonian is constructed are eventually given 
in Appendix \ref{sec:ss}.

\section{Notation}
\label{sec:notation}

Throughout this paper, we use the signature $(-,+,+,+)$ for the
metric. Spacetime tensor indices (ranging from 0 to 3) are denoted
with Greek letters, while spatial tensor indices (ranging from 1 to 3)
are denoted with lowercase Latin letters. Unless stated otherwise,
we use geometric units ($G=c=1$), although we restore the
factors of $c$ when expanding in PN orders.

We define a tetrad field as a set consisting of a timelike
future-oriented vector $\tilde{e}^\mu_{ T}$ and three spacelike
vectors $\tilde{e}^\mu_{ I}$ (${ I}=1,...,3$) --- collectively denoted
as $\tilde{e}^\mu_{ A}$ (${ A}=0,...,3$) --- satisfying
\begin{equation}\label{tetrad_orthonormal}
\tilde{e}^\mu_{ A}\,\tilde{e}^\nu_{ B}\, g_{\mu\nu} = \eta_{ AB}\,,
\end{equation}
where $\eta_{ TT}=-1$, $\eta_{ TI}=0$, $\eta_{ IJ}=\delta_{ IJ}$
($\delta_{ IJ}$ being the Kronecker symbol). 

Internal tetrad indices denoted with the uppercase Latin letters ${ A}$,
${ B}$, ${ C}$ and ${ D}$ always run from $0$ to $3$, while internal tetrad indices
with the uppercase Latin letters ${ I}$, ${ J}$, ${ K}$ and ${ L}$, associated
with the spacelike tetrad vectors, run from $1$ to $3$ only. The timelike
tetrad index is denoted by $T$.

Tetrad indices are raised and lowered with the metric $\eta_{ AB}$
[e.g., $\tilde{e}^\mu_{ A}=\eta_{ AB}\,(\tilde{e}^{ B})^\mu$]. 
We denote the projections of a vector $\boldsymbol{V}$ onto the
tetrad with $V^{ A}\equiv V^\mu\, \tilde{e}^{ A}_{\mu}$, and similarly for tensors
of higher rank. Partial derivatives  will be denoted with a comma or with
$\partial$, and covariant derivatives with a semicolon.

\section{Hamiltonian for a spinning test-particle in axisymmetric stationary spacetimes}
\label{sec:Hamiltonian_axisymmetric}

Following Ref.~\cite{bardeen_astres_occlus}, we write a generic axisymmetric stationary metric in quasi-isotropic 
coordinates as
\begin{eqnarray}
ds^2 = && -e^{2\nu}dt^2 +
R^2 \sin^2\theta B^2 e^{-2\nu} \left(d\phi -
\omega dt\right)^2 \nonumber \\
&& + e^{2\mu}\left(dR^2+R^2d\theta^2
\right)\;,
\label{eq:metric}
\end{eqnarray}
where $\nu$, $\mu$, $B$ and $\omega$ are functions of the coordinates $R$ and $\theta$. 
Introducing the cartesian quasi-isotropic coordinates
\begin{subequations}
\begin{eqnarray}
X&=&R \sin\theta \cos\phi\,,\\
Y&=&R \sin\theta \sin\phi\,,\\
Z&=&R \cos\theta\,,
\end{eqnarray}
\end{subequations}
we can write Eq.~(\ref{eq:metric}) as
\begin{eqnarray}
ds^2 &=& e^{-2 \nu}\,\left[B^2\,\omega^2 \left (X^2+Y^2\right)-e^{4 \nu}\right] dt^2 \nonumber \\
&& + 2 B^2\, e^{-2 \nu}\, \omega\,(Y\, dX - X\, dY)\,dt \nonumber \\
&& - 2 \frac{ \left(B^2 \,e^{-2 \nu}-e^{2 \mu}\right) X\, Y}{X^2+Y^2} dX\, dY \nonumber \\
&& +\frac{e^{2 \mu} \,X^2+B^2\, e^{-2 \nu} \,Y^2}{X^2+Y^2}\,dX^2 \nonumber \\
&& + \frac{B^2\, e^{-2 \nu}\, X^2+e^{2 \mu} \,Y^2}{X^2+Y^2}\,dY^2 + e^{2 \mu}\, dZ^2\,.\nonumber \\
\label{eq:metric_cartesian}
\end{eqnarray} 
It is straightforward to see that in the flat-spacetime limit ($\omega=\nu=\mu=0$, $B=1$)
Eq.~(\ref{eq:metric_cartesian}) reduces to the Minkowski metric. 

Reference~\cite{diracbrackets_ours} 
computed the Hamiltonian of a spinning test-particle in curved spacetime at linear order in the particle's spin, and showed that 
it can be written as 
\begin{equation}
H = H_{\rm NS} + H_{\rm S}\,,
\end{equation}
where ${H}_{\rm NS}$ is the Hamiltonian for a non-spinning test particle of mass $m$, given by
\begin{equation}\label{eq:Hns}
{H}_{\rm NS} = \beta^iP_i + \alpha \sqrt{m^2 + \gamma^{ij}\,P_i\,P_j}\,,
\end{equation} 
with
\begin{eqnarray}
\label{alphai}
\alpha &=& \frac{1}{\sqrt{-g^{tt}}}\,,\\
\beta^i &=& \frac{g^{ti}}{g^{tt}}\,,\\
\gamma^{ij} &=& g^{ij}-\frac{g^{ti} g^{tj}}{g^{tt}}\,,
\label{gammai}
\end{eqnarray}
and 
\begin{equation}
\label{Hamiltonian_canonical_final2}
{H}_{\rm S} = - \left(\beta^i\,F_i^K + F_t^K + \frac{\alpha \gamma^{ij}\,P_i\,F_j^K}{\sqrt{m^2 + 
\gamma^{ij}P_iP_j}}\right)\,S_K \,,\nonumber \\ 
\end{equation}
where the coefficients $F_\mu^I$ can be expressed in terms of a reference tetrad field $\tilde{e}_A$ as
\begin{eqnarray}\label{FmuK_def}
F_\mu^K &=& \left(2E_{\mu TI}\,\frac{\bar{\omega}_J}{\bar{\omega}_T} + E_{\mu IJ}\right)\,\epsilon^{IJK}\,,\\
E_{\lambda\mu\nu} &\equiv&  \frac{1}{2}\,\eta_{AB}\,\tilde{e}_\mu^A\,\tilde{e}_{\nu;\lambda}^{B}\,,
\end{eqnarray}
with 
\begin{eqnarray}
\bar{\omega}_\mu&=&\bar{P}_\mu-m \,\tilde{e}^{T}_\mu\,,\\
\bar{P}_i &=& P_i\,,\\
\bar{P}_t &=& -\beta^i\,P_i-\alpha\, \sqrt{m^2 +\gamma^{ij}\,P_i\, P_j}\,,\\
\label{bomegaT}
\bar{\omega}_T &=& \bar{\omega}_\mu\,\tilde{e}^\mu_{T}=\bar{P}_\mu\,\tilde{e}^\mu_{T}-m\,,\\
\bar{\omega}_I &=&\bar{\omega}_\mu\,\tilde{e}^\mu_{I}= \bar{P}_\mu\,\tilde{e}^\mu_{I}\,.
\label{bomegaI}
\end{eqnarray}
Reference~\cite{diracbrackets_ours} also showed that in order to obtain a Hamiltonian giving the usual leading-order spin-orbit 
coupling without gauge effects (or, equivalently, $H_{\rm S}=0$ in flat spacetime),  
the reference tetrad field must become {\it cartesian} in the flat-spacetime limit. We find that 
the following choice for the reference tetrad 
 \begin{subequations}
 \begin{align}
 &\tilde{e}^T_\alpha=\delta_\alpha^t(-g^{tt})^{-1/2}=e^\nu \,\delta_\alpha^t\,,\label{tetrad0}\\
 &\tilde{e}_1^\alpha=\frac{B\, e^{-\mu}\, X^2+e^\nu\, Y^2}{B \,(X^2+Y^2)} \,\delta^\alpha_X + \frac{\left(B\, e^{-\mu}-e^\nu\,\right) \,X \,Y}{B\,\left(X^2+Y^2\right)}\, \delta^\alpha_Y\nonumber \,,\\\label{tetrad1}\\
 &\tilde{e}_2^\alpha= \frac{\left(B \,e^{-\mu}-e^\nu\right)\, X\, Y}{B\, \left(X^2+Y^2\right)}\, \delta^\alpha_X +\frac{e^\nu\, X^2+
 B\, e^{-\mu} Y^2}{B \,(X^2 +Y^2)}\, \delta^\alpha_Y\nonumber \,,\\\label{tetrad2}\\
 &\tilde{e}_3^\alpha=e^{-\mu}\, \delta_Z^\alpha\label{tetrad3}\,,
 \end{align}
 \end{subequations}
%
%
indeed reduces to the cartesian tetrad $\tilde{e}^T_\alpha=1$, $\tilde{e}_I^\alpha=\delta_I^\alpha$
in the flat-spacetime limit.

We can then use the tetrad defined by Eqs.~(\ref{tetrad0})--(\ref{tetrad3}) to calculate 
the coefficients $F^K_\mu$ in Eq.~\eqref{FmuK_def}, and obtain 
\begin{equation}
\label{HS}
H_{\rm S} = H_{\rm SO} + H_{\rm SS}\,,
\end{equation}
with
\begin{widetext}
\begin{align} 
\label{HSO}
H_{\rm SO} =& \frac{e^{2 \nu -\mu } \left(e^{\mu +\nu }-B\right)
  (\boldsymbol{\hat{P}}\cdot \boldsymbol{\xi} R)\,S^Z}{B^2\, \sqrt{Q}\, R^2
  \,\xi^2}+ \frac{e^{\nu -2 \mu }}{B^2\, \left(\sqrt{Q}+1\right)\, \sqrt{Q}
  R^2\, \xi^2}\Bigg\{B_{\cos\theta} \,e^{\mu +\nu }
(\boldsymbol{\hat{P}}\cdot
\boldsymbol{\xi}\, R) \left(\sqrt{Q}+1\right)\, (\boldsymbol{S}\cdot \boldsymbol{N})\, \xi^2 \nonumber \\
& + R\, (\boldsymbol{S}\cdot \boldsymbol{\xi})\, \,\left[\mu_R
  (\boldsymbol{\hat{P}}\cdot \boldsymbol{V}\, R)
  \left(\sqrt{Q}+1\right)-\mu_{\cos\theta}\, (\boldsymbol{\hat{P}}\cdot
  \boldsymbol{N})\, \xi^2-\sqrt{Q}\, (\nu_R \,(\boldsymbol{\hat{P}}\cdot
  \boldsymbol{V}\, R)+(\mu_{\cos\theta}-\nu_{\cos\theta})
  (\boldsymbol{\hat{P}}\cdot \boldsymbol{N})\, \xi^2)\right]\, B^2 \nonumber \\
& + e^{\mu +\nu }\, (\boldsymbol{\hat{P}}\cdot \boldsymbol{\xi}\, R)
\left(2 \sqrt{Q}+1\right) \Big[\nu_R\, R\, (\boldsymbol{S}\cdot
\boldsymbol{V})-\nu_{\cos\theta}\, (\boldsymbol{S}\cdot \boldsymbol{N})
\,\xi^2\Big] \,B-B_R \,e^{\mu +\nu } (\boldsymbol{\hat{P}}\cdot
\boldsymbol{\xi}\, R)
\left(\sqrt{Q}+1\right)\, R\, (\boldsymbol{S}\cdot \boldsymbol{V})\Bigg\}\,,\nonumber \\
\end{align} 
\begin{align}
\label{HSS} 
H_{\rm SS} =& \omega\, S^Z+\frac{e^{-3 \mu -\nu }\, \omega_R}{2 B\, \left(\sqrt{Q}+1\right)\, \sqrt{Q}\, R
   \xi^2} \Bigg\{-e^{\mu +\nu }\, (\boldsymbol{\hat{P}}\cdot \boldsymbol{V}\, R) (\boldsymbol{\hat{P}}\cdot \boldsymbol{\xi}\, R) 
(\boldsymbol{S}\cdot \boldsymbol{\xi})\, B+e^{2 (\mu +\nu )}\, (\boldsymbol{\hat{P}}\cdot \boldsymbol{\xi}\, R)^2 (\boldsymbol{S}
\cdot \boldsymbol{V}) \nonumber \\
&+ e^{2 \mu }\, \left(1+\sqrt{Q}\right)\,\sqrt{Q}\, R^2\, (\boldsymbol{S}\cdot \boldsymbol{V})\, \xi^2\, B^2
+(\boldsymbol{\hat{P}}\cdot \boldsymbol{N})\, R \,\left[(\boldsymbol{\hat{P}}\cdot \boldsymbol{V} \,R) 
(\boldsymbol{S}\cdot \boldsymbol{N})
-(\boldsymbol{\hat{P}}\cdot \boldsymbol{N})\, R\, (\boldsymbol{S}\cdot \boldsymbol{V})\right]
   \xi^2\, B^2\,\Bigg\} \nonumber \\
& +\frac{e^{-3 \mu -\nu }\, \omega_{\cos\theta}}{2 B\, \left(\sqrt{Q}+1\right)
   \,\sqrt{Q}\, R^2} \Bigg\{e^{\mu +\nu }\, (\boldsymbol{\hat{P}}\cdot \boldsymbol{N}) 
(\boldsymbol{\hat{P}}\cdot \boldsymbol{\xi}\, R)\, R\, (\boldsymbol{S}\cdot \boldsymbol{\xi})\, B-e^{2 (\mu +\nu )} 
(\boldsymbol{\hat{P}}\cdot \boldsymbol{\xi}\, R)^2 (\boldsymbol{S}\cdot \boldsymbol{N}) \nonumber \\
& + \left[(\boldsymbol{S}\cdot \boldsymbol{N}) 
(\boldsymbol{\hat{P}}\cdot \boldsymbol{V}\, R)^2\!-\!(\boldsymbol{\hat{P}}\cdot \boldsymbol{N})\, R\,
 (\boldsymbol{S}\cdot \boldsymbol{V}) 
(\boldsymbol{\hat{P}}\cdot \boldsymbol{V} \,R)-e^{2 \mu }\,\left(1+\sqrt{Q}\right)\!\sqrt{Q}\, R^2
(\boldsymbol{S}\cdot \boldsymbol{N})\, \xi^2\right]\,B^2 \Bigg\}\,,
\end{align} 
\begin{equation}\label{eq:Q_general}
Q=1 + \gamma^{ij} {\hat{P}_i\,\hat{P}_j} = 1 + e^{-2 \mu }\, (\boldsymbol{\hat{P}}\cdot \boldsymbol{N})^2
+\frac{e^{-2 \mu }\, (\boldsymbol{\hat{P}}\cdot \boldsymbol{V}\, R)^2}{R^2 \,\xi^2}+\frac{e^{2 \nu } 
\,(\boldsymbol{\hat{P}}\cdot \boldsymbol{\xi}\, R)^2}{B^2\, R^2\,\xi^2}\,,
\end{equation}
\end{widetext}
where we denote
\begin{eqnarray} 
\hat{\boldsymbol{P}}&=&\frac{\boldsymbol{P}}{m}\,, \\
\boldsymbol{N}&=&\frac{\boldsymbol{X}}{R}\,, \\
\boldsymbol{\xi}&=&\boldsymbol{e}_Z\times\boldsymbol{N}=\frac{-Y\,\boldsymbol{e}_X + X\,\boldsymbol{e}_Y}{R}\,,\\
\boldsymbol{V}&=&\boldsymbol{N} \times \boldsymbol{\xi}\,,
\end{eqnarray} 
and 
\begin{eqnarray} 
f_R &\equiv& \frac{\partial f(R,\cos\theta)}{\partial R}\,,\\
f_{\cos\theta}&\equiv& \frac{\partial f(R,\cos\theta)}{\partial (\cos\theta)}\,.
\end{eqnarray} 
Here, the generic function $f$ can stand for $B$, $\omega$, $\mu$ or $\nu$.
Note that because $\omega$ is proportional to $g_{t\phi}$ [see Eq.~(\ref{eq:metric})]
and thus to the spin of the spacetime, $H_{\rm SS}$ (which is proportional to 
$\omega$ and its derivatives) gives the leading-order coupling between the particle's spin 
and the spin of the background spacetime
(together with other higher order terms). Also, because $\boldsymbol{\hat{P}}\cdot \boldsymbol{\xi}\, R=\hat{P}_\phi$
in spherical coordinates, $H_{\rm SO}$ is the part of the Hamiltonian which gives the leading-order spin orbit coupling (again, 
together with other higher order terms).
Moreover, note that $H_{\rm S}=0$ 
in a flat spacetime, thus confirming the absence of gauge effects in the leading order spin-orbit coupling.

As a consistency test, we specialize to the case of a spherically symmetric spacetime 
in quasi-isotropic coordinates, which was considered in Ref.~\cite{diracbrackets_ours} (see Sec.~V A 
therein). Because the metric for such a spacetime is given by 
\begin{equation}
ds^2 = -f(R) \,dt^2 + h(R)\,(dX^2+dY^2+dZ^2)\,,
\label{metricS}
\end{equation}
a comparison with Eq.~\eqref{eq:metric} immediately reveals that 
\begin{eqnarray} 
\label{BS}
B&=&\sqrt{f(R) h(R)}\,,\\
\omega&=&0\,,\\
\nu&=&\frac12 \log[f(R)]\,,\\
\mu&=&\frac12 \log[h(R)]\,.
\label{muS}
\end{eqnarray} 
Inserting Eqs.~(\ref{BS})--(\ref{muS}) in Eqs.~(\ref{HS})--(\ref{eq:Q_general}), we find 
\begin{eqnarray} 
\label{Hschw_QI}
 H_{\rm S}&=& \frac{\boldsymbol{L}\cdot\boldsymbol{S}}
{2 m\,R\,\sqrt{f(R)}\,h(R)^2\,\sqrt{Q}\,(1+\sqrt{Q})} \times \nonumber \\
&& \left \{ \sqrt{Q}[f^\prime(R)\, h(R) - f(R)\,h^\prime(R)] - 
f(R)\,h^\prime(R) \right \}\,,\nonumber \\
\end{eqnarray} 
where
\begin{eqnarray}
Q &=& 1 + \frac{1}{h}\,\hat{\boldsymbol{P}}^2\,, \\
\boldsymbol{L} &=& \boldsymbol{X}\times \boldsymbol{P}\,,
\end{eqnarray}
in agreement with Eq.~(5.7) in Ref.~\cite{diracbrackets_ours}.

Let us now investigate how the Hamiltonian (\ref{HS}) is affected by a change
of the radial coordinate $R$. Denoting the new radial
coordinate  by $r = |\boldsymbol{x}|$ and defining 
\begin{equation}
J^{-1}\equiv \frac{d R }{d r}\,,
\end{equation}
the radial derivatives of the metric potentials can be re-expressed as
\begin{equation}
f_R = {f_r}\,J \,,
\end{equation}
where again $f=B,\,\omega,\,\nu,\,\mu$. 
The spin $\boldsymbol{S}$, the derivatives of the metric potentials with respect to $\cos\theta$,
and the quantities
\begin{eqnarray}
\boldsymbol{N} &=& \boldsymbol{n} = \frac{\boldsymbol{x}}{r}\,,\\
\boldsymbol{\xi} &=& \boldsymbol{e}_Z\times\boldsymbol{N} = \boldsymbol{e}_z\times\boldsymbol{n}\,,\label{eq:xidef}\\
\boldsymbol{V} &=& \boldsymbol{v} = \boldsymbol{n} \times \boldsymbol{\xi}\,\\
&&\nonumber
\end{eqnarray}
are not affected by the coordinate change. The same applies to the quantities $\boldsymbol{\hat{P}}\cdot \boldsymbol{V}\, R$ and
$\boldsymbol{\hat{P}}\cdot \boldsymbol{\xi} R$ appearing in Eqs.~(\ref{HSO}), (\ref{HSS}) and~\eqref{eq:Q_general}. 
In fact, in spherical coordinates, we have
$\boldsymbol{\hat{P}}\cdot
\boldsymbol{V} R = -\hat{P}_\theta \sin\theta$ and
$\boldsymbol{\hat{P}}\cdot \boldsymbol{\xi} R = \hat{P}_\phi$, hence
\begin{eqnarray}
\boldsymbol{\hat{P}}\cdot \boldsymbol{V} R &=& \boldsymbol{\hat{p}}\cdot \boldsymbol{v} \,r\,,\\
\boldsymbol{\hat{P}}\cdot \boldsymbol{\xi} R &=& \boldsymbol{\hat{p}}\cdot \boldsymbol{\xi} \,r\,,
\end{eqnarray} 
where $\boldsymbol{\hat{p}} = \boldsymbol{p}/m$ and $\boldsymbol{p}$ is the conjugate momentum 
in the new coordinate system, i.e.,  $p_i=\partial X^{j}/\partial x^i\, P_j$. 
On the contrary, since $\boldsymbol{\hat{P}}\cdot \boldsymbol{N} =\hat{P}_R$, we have
\begin{equation}
\boldsymbol{\hat{P}}\cdot \boldsymbol{N} = {(\boldsymbol{\hat{p}}\cdot \boldsymbol{n})}\,{J}\,.
\end{equation} 
It is therefore straightforward to compute $H_{\rm S}$ 
in a coordinate system related to quasi-isotropic coordinates by a rescaling of the
radius. We have 
\begin{equation}
H_{\rm S} = H_{\rm SO} + H_{\rm SS}\,,
\label{Ht}
\end{equation}
where 
\begin{widetext}
\begin{eqnarray}
  H_{\rm SO} &=& \frac{e^{2 \nu -\mu }\, \left(e^{\mu +\nu }-B\right)\, (\boldsymbol{\hat{p}}\cdot \boldsymbol{\xi}\, r) \,S^z}
{B^2\,\sqrt{Q}\, R^2\, \xi^2}+
  \frac{e^{\nu -2 \mu }}{B^2 \,\left(\sqrt{Q}+1\right)\, \sqrt{Q}\, R^2\, \xi^2}
\Bigg\{B_{\cos\theta}\, e^{\mu +\nu } (\boldsymbol{\hat{p}}\cdot 
  \boldsymbol{\xi}\, r) \left(\sqrt{Q}+1\right) (\boldsymbol{S}\cdot \boldsymbol{n})\, \xi^2 \nonumber \\
  && + R\, (\boldsymbol{S}\cdot \boldsymbol{\xi}) J
  \left[\mu_r\, (\boldsymbol{\hat{p}}\cdot \boldsymbol{v}\, r) 
\left(\sqrt{Q}+1\right)-\mu_{\cos\theta} (\boldsymbol{\hat{p}}\cdot \boldsymbol{n}) \,
    \xi^2-\sqrt{Q}\, (\nu_r\, (\boldsymbol{\hat{p}}\cdot \boldsymbol{v}\, r)+(\mu_{\cos\theta}-\nu_{\cos\theta})
    (\boldsymbol{\hat{p}}\cdot \boldsymbol{n})\, \xi^2)\right]\, B^2 \nonumber \\
  && +e^{\mu +\nu }\, (\boldsymbol{\hat{p}}\cdot \boldsymbol{\xi}\, r) 
  \left(2 \sqrt{Q}+1\right) \Big[J\, \nu_r\, R\, (\boldsymbol{S}\cdot \boldsymbol{v})-\nu_{\cos\theta}
  \,(\boldsymbol{S}\cdot \boldsymbol{n})\, \xi^2\Big]\, B-J\, B_r\, e^{\mu +\nu }\, (\boldsymbol{\hat{p}}\cdot \boldsymbol{\xi}\, r) 
  \left(\sqrt{Q}+1\right)\, R\, (\boldsymbol{S}\cdot \boldsymbol{v})\Bigg\}\,, \nonumber \\\label{Ht1}
\end{eqnarray}
\begin{eqnarray}
H_{\rm SS} &=& \omega \,S^z+\frac{e^{-3 \mu -\nu }\, J\, \omega_r}{2 B\, \left(\sqrt{Q}+1\right)\, \sqrt{Q}\, R
   \,\xi^2} \Bigg\{ -e^{\mu +\nu }\, (\boldsymbol{\hat{p}}\cdot \boldsymbol{v}\, r) (\boldsymbol{\hat{p}}\cdot \boldsymbol{\xi}\, r) 
(\boldsymbol{S}\cdot \boldsymbol{\xi}) \,B+e^{2 (\mu +\nu )}\, (\boldsymbol{\hat{p}}\cdot \boldsymbol{\xi}\, r)^2 (\boldsymbol{S}
\cdot \boldsymbol{v}) \nonumber \\
&& + e^{2 \mu }\, \left(1+\sqrt{Q}\right)\,\sqrt{Q}\, R^2\, (\boldsymbol{S}\cdot \boldsymbol{v})\, \xi^2\, B^2\,
+J \,(\boldsymbol{\hat{p}}\cdot \boldsymbol{n}) \,R \left[(\boldsymbol{\hat{p}}\cdot \boldsymbol{v}\, r) (\boldsymbol{S}\cdot \boldsymbol{n})
-J\, (\boldsymbol{\hat{p}}\cdot \boldsymbol{n}) \,R (\boldsymbol{S}\cdot \boldsymbol{v})\right]
   \xi^2\, B^2\Bigg\} \nonumber \\
&& +\frac{e^{-3 \mu -\nu }\, \omega_{\cos\theta}}{2 B\, \left(\sqrt{Q}+1\right)
   \,\sqrt{Q}\, R^2} \Bigg\{
-e^{2 (\mu +\nu )}\, 
(\boldsymbol{\hat{p}}\cdot \boldsymbol{\xi}\, r)^2 \,(\boldsymbol{S}\cdot \boldsymbol{n}) \,
+e^{\mu +\nu }\, J\, (\boldsymbol{\hat{p}}\cdot \boldsymbol{n}) \,
(\boldsymbol{\hat{p}}\cdot \boldsymbol{\xi}\, r)\, R \,(\boldsymbol{S}\cdot \boldsymbol{\xi}) \,B\, \nonumber \\
&& + \left[(\boldsymbol{S}\cdot \boldsymbol{n}) \,
(\boldsymbol{\hat{p}}\cdot \boldsymbol{v} \,r)^2-J\, (\boldsymbol{\hat{p}}\cdot \boldsymbol{n})\, R\, 
(\boldsymbol{S}\cdot \boldsymbol{v}) \,(\boldsymbol{\hat{p}}\cdot \boldsymbol{v}\, r) -e^{2 \mu }\, 
\left(1+\sqrt{Q}\right)\,\sqrt{Q}\, R^2\,(\boldsymbol{S}\cdot \boldsymbol{n})\, \xi^2\right]\, B^2
\Bigg\}\,,\label{Ht2}
\end{eqnarray}
\begin{equation}
\label{Q}
Q=1 + \gamma^{ij}\, {\hat{p}_i\,\hat{p}_j} = 1 + e^{-2 \mu }\, (\boldsymbol{\hat{p}}\cdot \boldsymbol{n})^2 \,J^2
+\frac{e^{-2 \mu }\, (\boldsymbol{\hat{p}}\cdot \boldsymbol{v}\, r)^2}{R^2\, \xi^2}+\frac{e^{2 \nu } 
(\boldsymbol{\hat{p}}\cdot \boldsymbol{\xi}\, r)^2}{B^2\, R^2\,\xi^2}\,,
\end{equation}
\end{widetext}
and where $R$ must of course be expressed in terms of the new radial coordinate $r$.

\section{Hamiltonian for a spinning test-particle in Kerr spacetime in Boyer-Lindquist coordinates}
\label{sec:hamiltonian_kerr}

In this section, we will specialize the Hamiltonian derived in the previous section to the case of Kerr spacetime 
in Boyer-Lindquist coordinates.

We start from the metric potentials appearing in Eq.~(\ref{eq:metric}), which in the case of a Kerr spacetime take 
the form~\cite{cook_living_rev}
\begin{eqnarray}
\label{coeffB}
B&=&\frac{\sqrt{\Delta}}{R}\,,\\
\omega&=&\frac{2 a M r}{\Lambda}\,,\\
e^{2\nu}&=&\frac{\Delta\Sigma}{\Lambda}\,,\\
e^{2\mu}&=&\frac{\Sigma}{R^2}\,,
\label{coeffmu}
\end{eqnarray}
with
\begin{eqnarray}
\Sigma &=& r^2 + a^2\cos^2\theta \,, \\
\Delta &=& r^2 + a^2 - 2Mr \,, \\
\varpi^2 &=& r^2 + a^2 \,, \\
\Lambda &=& \varpi^4 - a^2\Delta\sin^2\theta \,,
\label{Lambda}
\end{eqnarray}
where the parameter $a$, which has the dimensions of a length, is related to the spin vector $\boldsymbol{S}_{\rm Kerr}$
of the Kerr black hole by
\eq
a= \frac{\vert\boldsymbol{S}_{\rm Kerr}\vert}{M}\,.
\eeq
The Boyer-Lindquist coordinate $r$ is related to the quasi-isotropic coordinate $R$ by
\begin{equation}
r=R + M + \frac{R_{\rm H}^2}{R}\,,
\end{equation}
where $R_{\rm H}=\sqrt{M^2 - a^2}/2$ is the horizon's radius in quasi-isotropic 
coordinates. Note that the inverse of this transformation is given, outside the horizon, by
\begin{equation}
R=\frac12 \big(r -M + \sqrt{\Delta}\big)\,.
\end{equation}
We then obtain that the derivatives of the metric potentials take the form
\begin{subequations}
\begin{eqnarray}
\label{eq1}
B_r &=& \frac{r - M - \sqrt{\Delta}}{R\,\sqrt{\Delta}}\,,\\
\omega_r &=& \frac{2 a\, M\, [\Sigma\, \varpi^2 - 2 r^2\, (\Sigma + \varpi^2)]}{\Lambda^2}\,,\\
\nu_r &=& \frac{r - M}{\Delta} + \frac{r}{\Sigma} \nonumber \\
&& - \frac{2 r\, \varpi^2 - a^2\, (r - M)\, \sin^2\theta}{\Lambda}\,,\\
\mu_r &=& \frac{r}{\Sigma} - \frac{1}{\sqrt{\Delta}}\,,\\
B_{\cos\theta} &=& 0\,,\\
\omega_{\cos\theta}&=& -\frac{4 a^3\, M\, r\, \Delta\, \cos\theta}{(\Delta\,\Sigma + 2 M \,r\, \varpi^2)^2}\,,\\
\nu_{\cos\theta}&=&\frac{2 a^2\, M\, r \,\varpi^2\, \cos\theta}{(\Delta\, \Sigma + 2 M\, r\, \varpi^2)\,\Sigma}\,,\\
\mu_{\cos\theta}&=&\frac{a^2\, \cos\theta}{\Sigma}\,,
\label{eq8}
\end{eqnarray}
\end{subequations}
and we also have
\begin{equation}
\label{eqJ}
J^{-1}= \frac{d R }{d r} = \frac{R}{\sqrt{\Delta}}\,.
\end{equation}
Inserting Eqs.~(\ref{eq1})--(\ref{eq8}) and Eq.~(\ref{eqJ}) into
Eqs.~(\ref{Ht})--(\ref{Q}), we find that $R$ cancels out both in
$Q$, that is
\begin{equation}
Q=1+\frac{\Delta (\boldsymbol{\hat{p}}\cdot \boldsymbol{n})^2}{\Sigma}+
\frac{(\boldsymbol{\hat{p}}\cdot \boldsymbol{\xi}\, r)^2 \,\Sigma }{\Lambda \,\sin^2\theta}+
\frac{(\boldsymbol{\hat{p}}\cdot \boldsymbol{v}\, r)^2 }{\Sigma\,\sin^2\theta}\,,
\end{equation}
and in the Hamiltonian $H_{\rm S}$. In conclusion, the Hamiltonian 
of a spinning test-particle in Kerr spacetime in Boyer-Lindquist coordinates is 
\begin{equation}
\label{Htotal}
H = H_{\rm NS} + H_{\rm S}\,,
\end{equation}
with 
\begin{equation}
\label{HNSKerr}
{H}_{\rm NS} = \beta^i\,p_i + \alpha\, \sqrt{m^2 + \gamma^{ij}\,p_i\,p_j}\,,
\end{equation} 
where $\alpha$, $\beta^i$ and $\gamma^{ij}$ are given in Eqs.~(\ref{alphai})--(\ref{gammai}) 
and need to be computed using the Kerr metric coefficients (\ref{coeffB})--(\ref{Lambda}), and with
\begin{equation}
\label{KerrHS}
H_{\rm S} = H_{\rm SO} + H_{\rm SS}\,,
\end{equation}
where
\begin{widetext}
\begin{eqnarray}
  H_{\rm SO} &=& \frac{e^{2 \nu -\tilde{\mu} }\,\left(e^{\tilde{\mu} +\nu }-\tilde{B}\right)\, 
    (\boldsymbol{\hat{p}}\cdot \boldsymbol{\xi}\, r)\,  (\boldsymbol{S}\cdot \boldsymbol{\hat{S}}_{\rm Kerr})}{\tilde{B}^2\, \sqrt{Q}\,\xi^2} + 
   \frac{e^{\nu -2 \tilde{\mu} }}{\tilde{B}^2\, \left(\sqrt{Q}+1\right)\, \sqrt{Q}\, \xi^2}\Bigg\{
  (\boldsymbol{S}\cdot \boldsymbol{\xi})\, \tilde{J}
  \left[\mu_r\, (\boldsymbol{\hat{p}}\cdot \boldsymbol{v}\, r) \left(\sqrt{Q}+1\right)-\mu_{\cos\theta}\, 
  (\boldsymbol{\hat{p}}\cdot \boldsymbol{n})\, 
    \xi^2 \right .\nonumber \\
  && \left . -\sqrt{Q}\, (\nu_r\, (\boldsymbol{\hat{p}}\cdot \boldsymbol{v}\, r)+(\mu_{\cos\theta}-\nu_{\cos\theta})\,
    (\boldsymbol{\hat{p}}\cdot \boldsymbol{n}) \,\xi^2)\right]\, \tilde{B}^2 +e^{\tilde{\mu} +\nu }\, 
  (\boldsymbol{\hat{p}}\cdot \boldsymbol{\xi}\, r)\, 
  \left(2 \sqrt{Q}+1\right)\, \Big[\tilde{J}\, \nu_r\, (\boldsymbol{S}\cdot \boldsymbol{v})-\nu_{\cos\theta}\,
  (\boldsymbol{S}\cdot \boldsymbol{n})\, \xi^2\Big]\, \tilde{B} \nonumber \\
  && -\tilde{J}\, \tilde{B}_r\, e^{\tilde{\mu} +\nu }\, (\boldsymbol{\hat{p}}\cdot \boldsymbol{\xi}\, r)\, 
  \left(\sqrt{Q}+1\right)\, (\boldsymbol{S}\cdot \boldsymbol{v})\Bigg\}\,,\label{KerrHSO}
\end{eqnarray}
and 
\begin{eqnarray}
H_{\rm SS} &=& \omega\, (\boldsymbol{S}\cdot \boldsymbol{\hat{S}}_{\rm Kerr})+\frac{e^{-3 \tilde{\mu} -\nu }\, \tilde{J}\, \omega_r}{2 \tilde{B}\, \left(\sqrt{Q}+1\right)\, \sqrt{Q}\, 
   \xi^2} \Bigg\{ -e^{\tilde{\mu} +\nu }\, (\boldsymbol{\hat{p}}\cdot \boldsymbol{v}\, r) \,
(\boldsymbol{\hat{p}}\cdot \boldsymbol{\xi}\, r)\, (\boldsymbol{S}\cdot \boldsymbol{\xi})\,
\tilde{B}+e^{2 (\tilde{\mu} +\nu )}\, (\boldsymbol{\hat{p}}\cdot \boldsymbol{\xi}\, r)^2\, (\boldsymbol{S}
\cdot \boldsymbol{v}) \nonumber \\
&& + e^{2 \tilde{\mu} }\, \left(1+\sqrt{Q}\right)\,\sqrt{Q}\,(\boldsymbol{S}\cdot \boldsymbol{v})\,\xi^2\, \tilde{B}^2
+\tilde{J} (\boldsymbol{\hat{p}}\cdot \boldsymbol{n})\, \left[(\boldsymbol{\hat{p}}\cdot \boldsymbol{v}\, r)\, 
(\boldsymbol{S}\cdot \boldsymbol{n}) -\tilde{J} (\boldsymbol{\hat{p}}\cdot \boldsymbol{n})\, (\boldsymbol{S}\cdot 
\boldsymbol{v})\right]\,\xi^2\,\tilde{B}^2\Bigg\} \nonumber \\
&& +\frac{e^{-3 \tilde{\mu} -\nu }\, \omega_{\cos\theta}}{2 \tilde{B}\, \left(\sqrt{Q}+1\right)\,
   \sqrt{Q} }\, \Bigg\{-e^{2 (\tilde{\mu} +\nu )}\, 
(\boldsymbol{\hat{p}}\cdot \boldsymbol{\xi} \,r)^2\, (\boldsymbol{S}\cdot \boldsymbol{n}) 
+ e^{\tilde{\mu} +\nu }\, \tilde{J}\, (\boldsymbol{\hat{p}}\cdot \boldsymbol{n})\, 
(\boldsymbol{\hat{p}}\cdot \boldsymbol{\xi}\, r)\, (\boldsymbol{S}\cdot \boldsymbol{\xi}) \tilde{B}
\nonumber \\
&& + \left[(\boldsymbol{S}\cdot \boldsymbol{n})\,(\boldsymbol{\hat{p}}\cdot \boldsymbol{v}\,r)^2-\tilde{J}\, 
(\boldsymbol{\hat{p}}\cdot \boldsymbol{n})\,(\boldsymbol{S}\cdot \boldsymbol{v})\, 
(\boldsymbol{\hat{p}}\cdot \boldsymbol{v}\,r)-e^{2 \tilde{\mu} }\,\left(1+\sqrt{Q}\right)\,\sqrt{Q}\, 
   (\boldsymbol{S}\cdot \boldsymbol{n}) \,\xi^2\right]\, \tilde{B}^2 
\Bigg\}\,,\label{KerrHSS}
\end{eqnarray}
\end{widetext}
where we define
\begin{eqnarray}
\tilde{B}&=&B \,R= \sqrt{\Delta}\,,\\
\tilde{B}_r&=&B_r\, R=\frac{r - M - \sqrt{\Delta}}{\sqrt{\Delta}}\,,\\
e^{2\tilde{\mu}} &=&e^{2\mu}\, R^2 = \Sigma\,,\\
\tilde{J}&=&J\, R = \sqrt{\Delta}\,,\\
\boldsymbol{\hat{S}}_{\rm Kerr}&=&\frac{\boldsymbol{{S}}_{\rm Kerr}}{\vert\boldsymbol{{S}}_{\rm Kerr}\vert}
\end{eqnarray}
and we recall that $\xi^2=\sin^2\theta$. We stress that because this Hamiltonian is expressed in terms
of quantities which are scalar under spatial rotations, we can express it in a cartesian coordinate system
in which the spin of the Kerr black hole is not directed along the $z$-axis. For that purpose, it is  
sufficient to replace $r$ with $(x^2+y^2+z^2)^{1/2}$, $\cos\theta$ with 
$\boldsymbol{\hat{S}}_{\rm Kerr}\cdot \boldsymbol{n}$, $\boldsymbol{e}_z$ with $\boldsymbol{\hat{S}}_{\rm Kerr}$ in Eq.~\eqref{eq:xidef}, and express the vectors appearing in Eqs.~(\ref{HNSKerr})--(\ref{KerrHSS}) 
in terms of their cartesian components.

As a consistency check, we can compute 
the Hamiltonian for a spinning test-particle in a Schwarzschild spacetime in Schwarzschild spherical coordinates by setting $a=0$,
and compare the result to the expression computed in Ref.~\cite{diracbrackets_ours} [see Eq.~(5.12) therein]. 
We find
\begin{align}
H_{\rm S}=&\frac{\psi^6}{R^3 \sqrt{Q}(1+\sqrt{Q})}\times\nonumber\\&\left[1-\frac{M}{2R} + 
2\left(1-\frac{M}{4R}\right)\sqrt{Q}\right](\boldsymbol{L}\cdot\boldsymbol{S}^\ast)\,,
\label{Hqi}
\end{align}
where 
\begin{eqnarray}
\boldsymbol{S}^\ast &=&\frac{M}{m} \,\boldsymbol{S}\,,\\
\psi &=& \left(1 + \frac{M}{2R}\right)^{-1}\,,\\
R &=& \frac12 \left(r-M+\sqrt{r^2-2 M r}\right)\,,
\end{eqnarray}
and
\begin{equation}
 Q=1+(\boldsymbol{\hat{p}}\cdot \boldsymbol{n})^2\, \left(1-\frac{2 M}{r}\right)
+\frac{(\boldsymbol{\hat{p}}\cdot \boldsymbol{v})^2+(\boldsymbol{\hat{p}}\cdot 
\boldsymbol{\xi})^2 }{\sin^2\theta}\,,
\end{equation}
in agreement with Ref.~\cite{diracbrackets_ours}.
Also, it is worth noting that the Hamiltonian (\ref{Hqi}) is the same as the quasi-isotropic Schwarzschild
Hamiltonian~\eqref{Hschw_QI}, expressed in terms of the Schwarzschild
coordinate $r$. This is because the scalar
product $\boldsymbol{L}\cdot\boldsymbol{S}$ is unaffected by a change
of the radial coordinate. 

\section{Effective-one-body Hamiltonian for two spinning black holes}
\label{sec:deformed_metric}

The EOB approach was originally introduced in Refs.~\cite{Buonanno99,Buonanno00,DJS3PN,
Damour01c} to provide us with an {\it improved} (resummed) Hamiltonian that could be used 
to evolve a binary system not only during the long inspiral, but also during 
the plunge, and that could supply a natural {\it moment} at which to switch 
from the two body description to the one-body description,
in which the system is represented by a superposition of quasi-normal 
modes of the remnant black hole. 

A crucial ingredient of the EOB approach 
is the {\it real} PN-expanded Arnowitt-Deser-Misner (ADM) Hamiltonian (or {\it real} 
Hamiltonian) describing two black holes of masses $m_1, m_2$ and spins 
$\boldsymbol{S}_1$, $\boldsymbol{S}_2$. The real Hamiltonian is then 
canonically transformed and subsequently \textit{mapped} to an {\it effective} Hamiltonian $H_{\rm eff}$
describing a test-particle of mass $\mu=m_1\,m_2/(m_1+m_2)$ and suitable
spin $\boldsymbol{S}^\ast$, moving in a \textit{deformed} Kerr metric of mass $M = m_1 + m_2$ and 
suitable spin $\boldsymbol{S}_{\rm Kerr}$. 
The parameter regulating the deformation is the symmetric mass ratio of the binary, $\eta=\mu/M$,
which ensures that the deformation disappears in the case of extreme mass-ratio binaries. The resulting improved EOB 
Hamiltonian then takes the form
\begin{equation}
\label{hreal0}
H^{\rm improved}_\mathrm{real} = M\,\sqrt{1+2\eta\,\left(\frac{H_{\rm eff}}{\mu}-1\right)}\,.
\end{equation}

The computation of the improved EOB Hamiltonian consists of several stages. For this reason,
we briefly review here the main steps and the underpinning logic that
we will follow in the rest of this section:
\begin{enumerate}
\item[(i)] We apply a canonical transformation to the PN-expanded ADM Hamiltonian using a generating function 
which is compatible with the one used in previous EOB work, obtaining the PN-expanded Hamiltonian 
in EOB canonical coordinates (see Sec.~\ref{sec:ADM});
\item[(ii)]  We compute the effective Hamiltonian corresponding to the canonically 
transformed PN-expanded ADM Hamiltonian (see Sec.~\ref{sec:eff});
\item[(iii)] We deform  the Hamiltonian of a spinning test-particle 
in Kerr derived in Sec.~\ref{sec:hamiltonian_kerr} 
by deforming the Kerr metric (see Sec.~\ref{sec:kerrdeformed}) , and expand this deformed Hamiltonian in PN orders (see Sec.~\ref{sec:pn});
\item[(iv)] Comparing (iii) and (iii), we work out the mapping between the 
spin variables in the real and effective descriptions, and write the improved EOB Hamiltonian (see Sec.~\ref{sec:EOB}). 
\end{enumerate}

\subsection{The ADM Hamiltonian canonically transformed to EOB coordinates}
\label{sec:ADM}

We denote the ADM canonical variables in the 
binary's center-of-mass frame with 
$\boldsymbol{r}^\prime$ and $\boldsymbol{p}^\prime$.
It is convenient to introduce the following spin combinations:
\begin{eqnarray}
\boldsymbol{\sigma}&=&\boldsymbol{S}_1+\boldsymbol{S}_2\,, \label{sigma}\\
\boldsymbol{\sigma}^\ast&=&\boldsymbol{S}_1\,\frac{m_2}{m_1}+\boldsymbol{S}_2\,\frac{m_1}{m_2}\,,
\label{sigmastar}\\
\boldsymbol{\sigma}_0&=& \boldsymbol{\sigma} + \boldsymbol{\sigma}^\ast\,.
\end{eqnarray}
Moreover, in order to consistently keep 
track of the PN orders, we will restore the speed of light $c$ and rescale the spins variables as 
$\boldsymbol{{\sigma}}^\ast \rightarrow \boldsymbol{\sigma}^\ast \,c$ and 
$\boldsymbol{{\sigma}} \rightarrow \boldsymbol{\sigma} \,c$.\footnote{This is appropriate for
 black holes or a rapidly rotating compact stars. In
  the black-hole case, $S=\chi M^2/c$, with $\chi$ ranging from $0$ to $1$. 
In the rapidly spinning star case one has $S=M v_{\rm rot} r\sim
  M c r_{s}\sim  M^2/c$ (where we have assumed that the rotational
  velocity $v_{\rm rot}$ is comparable to $c$ and that the stellar
  radius $r$ is of the order of the Schwarzschild radius
  $r_s\sim M/c^2$).}
The canonical ADM Hamiltonian is known through 3PN 
order~\cite{Damour-Schafer:1988,Damour:2007nc,SHS07,SSH08,SHS08} 
and partially at higher PN orders \cite{Hergt:2007ha,hergt_schafer_08}. In particular, the spin-orbit and spin-spin 
coupling terms agree with those  
computed via effective-field-theory techniques at 1.5PN, 2PN and 3PN order~\cite{PR06,PR07,PR08b,PR08a}.
In this paper, we use the spin-independent part of the ADM Hamiltonian through 3PN order, but we
only use its spin-dependent part through 2.5 PN order, 
i.e., we consider the leading-order (1.5 PN) and the next-to-leading order (2.5PN) spin-orbit couplings, 
but only the leading order (2PN) spin-spin coupling. The expressions for these 
couplings are~\cite{Damour01c,Damour:2007nc}
\begin{eqnarray}
\label{HSO_adm}
 H^{\rm ADM}_{\rm SO}(\boldsymbol{r}^{\prime},\boldsymbol{p}^{\prime},\boldsymbol{\sigma}^\ast,\boldsymbol{\sigma}) &=& 
\frac{1}{c^3}\,\frac{\boldsymbol{L}^\prime}{r^{\prime\,3}}\cdot (g^{\rm ADM}_\sigma \,
\boldsymbol{\sigma}+ g^{\rm ADM}_{\sigma^\ast}\,\boldsymbol{\sigma}^\ast)\,, \nonumber \\ \\\label{HSS_adm}
H^{\rm ADM}_{\rm SS}(\boldsymbol{r}^{\prime},\boldsymbol{p}^{\prime},\boldsymbol{\sigma}^\ast,\boldsymbol{\sigma}) 
&=& \frac{1}{c^4}\,\frac{\eta}{2 r^{\prime\,3}}\,\left [3(\boldsymbol{n}^\prime \cdot \boldsymbol{\sigma}_0)^2 - 
 \boldsymbol{\sigma}_0^2 \right ]\,,\nonumber \\
\end{eqnarray}
with $\boldsymbol{L}^\prime = \boldsymbol{r}^\prime \times \boldsymbol{p}^\prime$, 
$\boldsymbol{n}^\prime= \boldsymbol{r}^\prime/r^\prime$, and
\begin{subequations}
\begin{eqnarray}
g^{\rm ADM}_{\sigma} &=& 2 + \frac{1}{c^2} \left [\frac{19}{8}\, \eta\,\boldsymbol{\hat{p}}^{\prime \,2} + 
\frac{3}{2} \eta\, (\boldsymbol{n}^{\prime}\cdot \boldsymbol{\hat{p}}^\prime)^2 \right . \nonumber \\
&& \left. - (6 + 2\eta)\, \frac{M}{r^{\prime}} \right ]\,, \\
g^{\rm ADM}_{\sigma^\ast} &=& \frac{3}{2} + \frac{1}{c^2} \left [ 
\left (-\frac{5}{8} + 2 \eta \right )\,\boldsymbol{\hat{p}}^{\prime\,2} + \frac{3}{4} \eta \,
(\boldsymbol{n}^{\prime} \cdot \boldsymbol{\hat{p}}^\prime)^2 \right . \nonumber \\
&& \left . - (5 + 2 \eta) \frac{M}{r^{\prime}} \right ]\,,
\end{eqnarray}
\end{subequations}
where we have introduced the rescaled conjugate momentum
$\boldsymbol{\hat{p}}^\prime = \boldsymbol{p}^\prime/\mu$.

We now perform a canonical transformation from the ADM canonical
variables $\boldsymbol{r^\prime}$ and $\boldsymbol{p^\prime}$ to the
EOB canonical variables $\boldsymbol{r}$ and $\boldsymbol{p}$.
Let us first consider the purely orbital generating function
\begin{align}
G(\boldsymbol{r}^\prime,\boldsymbol{p})=&\,\boldsymbol{r}^\prime\cdot
\boldsymbol{p}+G_{\rm NS}(\boldsymbol{r}^\prime,\boldsymbol{p})\,,\\
G_{\rm NS}(\boldsymbol{r}^\prime,\boldsymbol{p})=&\,G_{\rm NS\, 1PN}(\boldsymbol{r}^\prime,\boldsymbol{p})\nonumber\\&
+G_{\rm NS\, 2PN}(\boldsymbol{r}^\prime,\boldsymbol{p})
+G_{\rm NS\, 3PN}(\boldsymbol{r}^\prime,\boldsymbol{p})\,,
\end{align}
where the 1PN-accurate generating function $G_{\rm NS\, 1PN}$ was derived in Ref.~\cite{Buonanno99},
\begin{equation}
G_{\rm NS\, 1PN}(\boldsymbol{r}^\prime,\boldsymbol{p})=\frac{1}{c^2} 
\boldsymbol{r^\prime}\cdot\boldsymbol{p}\,\left[-\frac12 \eta\,\boldsymbol{{\hat{p}}}^{2} 
+ \frac{M}{r^\prime}\left(1+\frac12\eta\right)\,\right]\,,
\end{equation}
while the 2PN and 3PN accurate generating functions, $G_{\rm NS\, 2PN}$ and $G_{\rm NS\, 3PN}$,
were derived in Refs.~\cite{Buonanno99} and \cite{DJS3PN}, respectively.
From the definition of generating function, it follows that the
transformation of the phase-space variables is implicitly given by
\begin{gather}\label{x_transf}
x^i=x^{\prime i}+\frac{\partial G_{\rm NS}(x',p)}{\partial p_i}\,,\\
p_i=p^{\prime}_{i}-\frac{\partial G_{\rm NS}(x',p)}{\partial x^{\prime i}}\label{p_transf}\,,
\end{gather}
while the Hamiltonian transforms as $H(\boldsymbol{r},\boldsymbol{p})=H^{\rm ADM}(\boldsymbol{r}^\prime,\boldsymbol{p}^\prime)$.
At linear order, which is enough for our purposes, Eqs.~\eqref{x_transf} and \eqref{p_transf}
can be written as $y=y'-\{G_{\rm NS},y'\}$, where $\{...\}$ are the Poisson brackets and where $y$ stands for either $x$ or $p$. The transformation
of the Hamiltonian, again at linear order, is then $H(y)=H^{\rm ADM}(y)+\{G_{\rm NS},H^{\rm ADM}\}(y)$~\cite{DJS08}.
Similarly, if one considers a generating
function which depends not only on the orbital variables, but also on the spins,
\begin{multline} \label{G}
G(\boldsymbol{r}^\prime,\boldsymbol{p},\boldsymbol{\sigma}^\ast,\boldsymbol{\sigma})=\boldsymbol{r}^\prime\cdot
\boldsymbol{p}\\+G_{\rm NS}(\boldsymbol{r}^\prime,\boldsymbol{p})+G_{\rm S}(\boldsymbol{r}^\prime,\boldsymbol{p},\boldsymbol{\sigma}^\ast,\boldsymbol{\sigma})
\end{multline}
the Hamiltonian will again transform as $H(y)=H^{\rm ADM}(y)+\{G_{\rm NS},H^{\rm ADM}\}(y)+\{G_{\rm S},H^{\rm ADM}\}(y)$, 
where now the Poisson brackets in the term $\{G_{\rm S},H^{\rm ADM}\}$ will involve also the spin variables~\cite{DJS08}.
In particular, let us consider a spin-dependent generating function 
\begin{align}
G_{\rm S}(\boldsymbol{r}^\prime,\boldsymbol{p},\boldsymbol{\sigma}^\ast,\boldsymbol{\sigma})=&\,G_{\rm S\, 2PN}(\boldsymbol{r}^\prime,\boldsymbol{p},\boldsymbol{\sigma}) 
  \nonumber \\&+G_{\rm S\, 2.5PN}(\boldsymbol{r}^\prime,\boldsymbol{p},
\boldsymbol{\sigma}^\ast,\boldsymbol{\sigma})
\\&  + 
G_{\rm SSS\, 2.5PN}(\boldsymbol{r}^\prime,\boldsymbol{p},\boldsymbol{\sigma}^\ast,\boldsymbol{\sigma})
\nonumber \,.
\end{align}
where
%
the 2PN-accurate spin-dependent generating function $G_{\rm S\, 2PN}$ was implicitly\footnote{See discussion in Sec II D of Ref.~\cite{Damour01c}. The need for this generating function will become apparent with Eq.~\eqref{eq:HNSSS} in Sec.~\ref{sec:pn}.} used in Ref.~\cite{Damour01c},
\begin{align}
G_{\rm S\, 2PN}(\boldsymbol{r}^\prime,\boldsymbol{p},\boldsymbol{\sigma}) =&\;
 -\frac{1}{2 c^4 M^2 r^{\prime2}}\,\Big\{ [\boldsymbol{\sigma}^2 - (\boldsymbol{\sigma}\cdot \boldsymbol{n}^\prime)^2]
(\boldsymbol{r}^\prime\cdot\boldsymbol{p})\nonumber \\&+(\boldsymbol{\sigma}\cdot \boldsymbol{n}^\prime) 
(\boldsymbol{r}^\prime\times\boldsymbol{p})\cdot (\boldsymbol{\sigma}\times \boldsymbol{n}^\prime)\Big\}\,;
\end{align}
the 2.5PN-accurate generating function $G_{\rm S\, 2.5PN}$ linear in the spin variables was introduced in Ref.~\cite{DJS08},
\begin{eqnarray}
G_{\rm S\, 2.5PN}(\boldsymbol{r}^\prime,\boldsymbol{p},\boldsymbol{\sigma}^\ast,\boldsymbol{\sigma})&=&
\frac{1}{\mu\, r^{\prime 3}\, c^5}\,
(\boldsymbol{r}^\prime\cdot\boldsymbol{p})(\boldsymbol{r}^\prime \times \boldsymbol{p}) 
\cdot \nonumber \\
&& \left[a(\eta)\, \boldsymbol{\sigma} +b(\eta)\, \boldsymbol{\sigma}^\ast\right]\,,
\end{eqnarray}
$a(\eta)$ and $b(\eta)$ being arbitrary gauge functions; also,
for reasons which will become clear in Sec.~\ref{sec:pn}, we include the 
following 2.5PN-accurate generating function, cubic in the spins,
\begin{equation}
G_{\rm SSS\, 2.5PN}(\boldsymbol{r}^\prime,\boldsymbol{p},\boldsymbol{\sigma}^\ast,\boldsymbol{\sigma})=
\frac{\mu}{2 M^3 r^{\prime\,4} c^5} (\boldsymbol{\sigma} \cdot \boldsymbol{r}^\prime) 
[\boldsymbol{\sigma}^\ast\cdot (\boldsymbol{\sigma} \times \boldsymbol{r^\prime})]\,.
\label{G25PNnew}
\end{equation}

When applying the generating function (\ref{G}) to the ADM 2PN spin-spin Hamiltonian~\eqref{HSS_adm},
we obtain
\begin{eqnarray}
H_{\rm SS\, 2PN}(\boldsymbol{r},\boldsymbol{p},
\boldsymbol{\sigma}^\ast,\boldsymbol{\sigma}) &=& H^{\rm ADM}_{\rm SS\, 2PN}(\boldsymbol{r},
\boldsymbol{p},
\boldsymbol{\sigma}^\ast,\boldsymbol{\sigma})\nonumber \\
&& \!\!\!\!\! +\{G_{\rm S\, 2PN}, H_{\rm Newt}\}(\boldsymbol{r},\boldsymbol{p},
\boldsymbol{\sigma})\,,\nonumber \\
\end{eqnarray}
with
\begin{align}
 &H_{\rm Newt}=-\frac{M\,\mu}{r}+\frac{\boldsymbol{p}^2}{2\mu}\,,
\\&\nonumber\\
&\{G_{\rm S\, 2PN}, H_{\rm Newt}\}(\boldsymbol{r},\boldsymbol{p},
\boldsymbol{\sigma})=\nonumber
-\frac{1}{c^4}\,\frac{\eta}{2 r^{3}}\,\left[(\boldsymbol{n} \cdot \boldsymbol{\sigma})^2 - 
 \boldsymbol{\sigma}^2\right]\\ &\qquad\qquad\qquad+ \frac{1}{2 \mu\, M^2\, r^2\, c^4}\Big\{-[\boldsymbol{p}^2 -
 2 (\boldsymbol{p}\cdot \boldsymbol{n})^2] \boldsymbol{\sigma}^2 \nonumber \\&\qquad\qquad\qquad+
[(\boldsymbol{p}-2 (\boldsymbol{p}\cdot \boldsymbol{n})\boldsymbol{n})\cdot\boldsymbol{\sigma}]\, 
\boldsymbol{p}\cdot\boldsymbol{\sigma}\Big\}\,.
\end{align}
Similarly, if we apply the same generating function to the ADM spin-orbit Hamiltonian 
(\ref{HSO_adm}), the 1.5PN order term remains unaltered~\cite{DJS08}, 
while the 2.5PN order term transforms as~\cite{DJS08}
\begin{align}
H_{\rm SO\, 2.5PN}&\,(\boldsymbol{r},\boldsymbol{p},
\boldsymbol{\sigma}^\ast,\boldsymbol{\sigma}) = H^{\rm ADM}_{\rm SO\, 2.5PN}(\boldsymbol{r},
\boldsymbol{p},
\boldsymbol{\sigma}^\ast,\boldsymbol{\sigma})\nonumber \\
& +\{G_{\rm 2.5PN}, H_{\rm Newt}\}(\boldsymbol{r},\boldsymbol{p},
\boldsymbol{\sigma}^\ast,\boldsymbol{\sigma})\,,\nonumber \\
 &+\{G_{\rm NS\,1PN}, H^{\rm ADM}_{\rm SO\, 1.5PN}\}(\boldsymbol{r},\boldsymbol{p},
\boldsymbol{\sigma}^\ast,\boldsymbol{\sigma})
\end{align}
where
\begin{equation}
G_{\rm 2.5PN} = G_{\rm SS\, 2.5PN} + G_{\rm SSS\, 2.5PN}\,, 
\end{equation}
\begin{eqnarray}
&& \{G_{\rm 2.5PN}, H_{\rm Newt}\}(\boldsymbol{r},\boldsymbol{p},
\boldsymbol{\sigma}^\ast,\boldsymbol{\sigma}) = \nonumber \\
&& \frac{1}{r^{3} c^5}\,
\boldsymbol{L} \cdot \left [b(\eta)\, \boldsymbol{\sigma^\ast} 
+ a(\eta)\, \boldsymbol{\sigma}\right]\,
\left[-\frac{M}{r} + \boldsymbol{\hat{p}}^{2} - 3 (\boldsymbol{\hat{p}} \cdot \boldsymbol{n})^2 \right]
\nonumber\\
&&  + \frac{[\boldsymbol{\sigma}^\ast\cdot (\boldsymbol{\sigma}\times 
\boldsymbol{n})][\boldsymbol{\sigma}\cdot(\boldsymbol{p} -2 (\boldsymbol{p}
\cdot \boldsymbol{n})\, 
\boldsymbol{n})]}{M^3\,r^{3}\, c^5} \nonumber\\
&&  +\frac{(\boldsymbol{L} \cdot\boldsymbol{\sigma}^\ast) \boldsymbol{\sigma}^2
-(\boldsymbol{L} \cdot\boldsymbol{\sigma}) 
(\boldsymbol{\sigma}^\ast\cdot\boldsymbol{\sigma})}{2 M^3\, r^{4} c^5}\,,
\end{eqnarray}
and 
\begin{align}
&\{G_{\rm NS\,1PN}, H^{\rm ADM}_{\rm SO\, 1.5PN}\}(\boldsymbol{r},\boldsymbol{p},
\boldsymbol{\sigma}^\ast,\boldsymbol{\sigma})=\nonumber\\&
-\frac{3\boldsymbol{L}}{2r^{3} c^5}
\! \cdot\! \left(\frac32\boldsymbol{\sigma^\ast}\!+\!2\boldsymbol{\sigma}\!\right)
\left\{\!-\frac{M}{r} (2+\eta)  + \eta \left[\boldsymbol{\hat{p}}^{2} +2 (\boldsymbol{\hat{p}} \cdot \boldsymbol{n})^2\right]\! \right\}.
\nonumber\\&
\end{align}
Therefore, the complete real Hamiltonian in the EOB canonical coordinates is
\begin{eqnarray}
\label{HADPN}
H(\boldsymbol{r},\boldsymbol{p},
\boldsymbol{\sigma}^\ast,\boldsymbol{\sigma}) &=& H_{\rm nospin}(\boldsymbol{r},\boldsymbol{p},
\boldsymbol{\sigma}^\ast,\boldsymbol{\sigma}) \nonumber \\
&& + H^{\rm ADM}_{\rm SO}(\boldsymbol{r},\boldsymbol{p},
\boldsymbol{\sigma}^\ast,\boldsymbol{\sigma}) \nonumber \\
&& + H^{\rm ADM}_{\rm SS}(\boldsymbol{r},\boldsymbol{p},
\boldsymbol{\sigma}^\ast,\boldsymbol{\sigma}) \nonumber \\
&& + \{G_{\rm 2.5PN}, H_{\rm Newt}\}(\boldsymbol{r},\boldsymbol{p},
\boldsymbol{\sigma}^\ast,\boldsymbol{\sigma})\nonumber\\
&& +\{G_{\rm NS\,1PN}, H^{\rm ADM}_{\rm SO\, 1.5PN}\}(\boldsymbol{r},\boldsymbol{p},
\boldsymbol{\sigma}^\ast,\boldsymbol{\sigma})\nonumber\\
&&  +\{G_{\rm S\, 2PN}, H_{\rm Newt}\}(\boldsymbol{r},\boldsymbol{p},
\boldsymbol{\sigma})\,, \nonumber \\
\end{eqnarray}
where $H_{\rm nospin}$ is the 3PN ADM Hamiltonian for non-spinning black holes, 
canonically transformed to EOB coordinates, which can be obtained 
from Ref.~\cite{DJS3PN}.

\subsection{Spin couplings in the effective Hamiltonian}
\label{sec:eff}

Following Refs.~\cite{Buonanno00,DJS3PN,Damour01c}, we 
map the effective and real two-body Hamiltonians as
\begin{equation}
\label{heff}
\frac{H_{\rm eff}}{\mu c^2} =
\frac{H_{\rm real}^2 - m_1^2\,c^4 - m_2^2\, c^4}
{2 m_1\, m_2\, c^4}\,,
\end{equation}
where $H_{\rm real}$ is the real two-body Hamiltonian containing also the rest-mass contribution $M\,c^2$. 
We denote the non-relativistic part of the real Hamiltonian by $H^{\rm NR}$, i.e., 
$H^{\rm NR} \equiv H_{\rm real} - M\,c^2$. Identifying $H^{\rm NR}$ 
with $H$ as given in Eq.~(\ref{HADPN}), and 
expanding Eq.~(\ref{heff}) in powers of $1/c$, we find that 
the 1.5PN and 2.5PN order spin-orbit couplings of the effective 
Hamiltonian are
\begin{align}
\label{eq:HeffSO}
H^{\rm eff}_{\rm SO}(\boldsymbol{r},\boldsymbol{p},
\boldsymbol{\sigma}^\ast,\boldsymbol{\sigma})=&\frac{1}{c^3}\frac{\boldsymbol{L}}{r^3}\cdot
\left(g_{\sigma}^{\rm eff}\,\boldsymbol{\sigma}+g_{\sigma^\ast}^{\rm eff}\,\boldsymbol{\sigma}^{\ast}\right)\nonumber\\&+
\frac{[\boldsymbol{\sigma}^\ast\cdot (\boldsymbol{\sigma}\times 
\boldsymbol{n})][\boldsymbol{\sigma}\cdot(\boldsymbol{p}-2 (\boldsymbol{p}\cdot \boldsymbol{n})\, 
\boldsymbol{n})]}{M^3 r^3 c^5}\nonumber\\&+\frac{(\boldsymbol{L}\cdot\boldsymbol{\sigma}^\ast) \boldsymbol{\sigma}^2
-(\boldsymbol{L}\cdot\boldsymbol{\sigma}) 
(\boldsymbol{\sigma}^\ast\cdot\boldsymbol{\sigma})}{2 M^3 r^4 c^5}\,,
\end{align}
where~\cite{DJS08}
\begin{subequations}
\begin{eqnarray}
g^{\rm eff}_\sigma&=& 2 + \frac{1}{c^2}\,
\left \{ \left [\frac{3}{8}\,\eta + a(\eta) \right ]\,\boldsymbol{\hat{p}}^2 \right. \nonumber \\
&& \left. - \left [\frac{9}{2}\,\eta + 3 \,a(\eta)\right ]\,(\boldsymbol{\hat{p}}\cdot\boldsymbol{n})^2 
\right . \nonumber\\
&& \left . - \frac{M}{r}\,\left [\eta + a(\eta) \right ] \right \}\,, \\
g^{\rm eff}_{\sigma^\ast} &=&\frac{3}{2} +
\frac{1}{c^2}\, \left \{ \left [-\frac{5}{8}+\frac{1}{2}\,\eta + b(\eta)\right ] \,
\boldsymbol{\hat{p}}^2 \right . \nonumber\\
&& \left . - \left [ \frac{15}{4}\,\eta+3\,b(\eta) \right ]\,(\boldsymbol{\hat{p}}\cdot\boldsymbol{n})^2 
\right. \nonumber\\
&& \left . - \frac{M}{r}\,\left [ \frac{1}{2}+\frac{5}{4}\,\eta+b(\eta)\right ] \right \}\,,
\end{eqnarray}
\end{subequations}
and the 2PN order spin-spin coupling is
\begin{align}
\label{eq:HeffSS}
&{H}_{\rm SS}^{\rm eff}(\boldsymbol{r},\boldsymbol{p},
\boldsymbol{\sigma}^\ast,\boldsymbol{\sigma}) = \frac{1}{c^4}\,\frac{\eta}{2 \,r^3}\,(3n_{i}\, n_j
 - \delta_{ij})\,\sigma_0^i\, \sigma_0^j \nonumber\\&\qquad\qquad\qquad\nonumber
-\frac{1}{c^4}\,\frac{\eta}{2 r^{3}}\,\left[(\boldsymbol{n} \cdot \boldsymbol{\sigma})^2 - 
 \boldsymbol{\sigma}^2\right]\\ &\qquad\qquad\qquad+ \frac{1}{2 \mu M^2\,r^2\, c^4}\Big\{-[\boldsymbol{p}^2 -
 2 (\boldsymbol{p}\cdot \boldsymbol{n})^2] \,\boldsymbol{\sigma}^2 \nonumber \\&\qquad\qquad\qquad+
[(\boldsymbol{p}-2 (\boldsymbol{p}\cdot \boldsymbol{n})\,\boldsymbol{n})\cdot\boldsymbol{\sigma}]\, 
\boldsymbol{p}\cdot\boldsymbol{\sigma}\Big\}\,.
\end{align}

\subsection{The Hamiltonian of a spinning test-particle in a deformed Kerr spacetime}
\label{sec:kerrdeformed}

We now deform the Hamiltonian of a spinning test-particle in a Kerr spacetime 
computed in Sec.~\ref{sec:hamiltonian_kerr} [see Eqs.~\eqref{HNSKerr}, 
\eqref{KerrHS}, \eqref{KerrHSO} and \eqref{KerrHSS}] by deforming the Kerr metric. 
The deformation that we introduce is regulated by the parameter $\eta=\mu/M$, and therefore disappears
in the test-particle limit.
Also, the deformed Hamiltonian will 
be such as to reproduce, when expanded in PN orders, the spin 
couplings of the effective Hamiltonian given in Sec.~\ref{sec:eff}.

When the spin of the Kerr black hole is zero, that is $a=0$, we require 
the metric to coincide with the deformed-Schwarzschild metric
 used in the EOB formalism for non-spinning black-hole
binaries~\cite{Buonanno99,DJS3PN}. That deformation 
simply amounts to changing the components $g_{tt}$ and $g_{rr}$ of the metric. 
In the spinning case, following Ref.~\cite{DJS08}, we seek an extension 
of this deformation by changing the potential $\Delta$ appearing 
in the Kerr potentials (\ref{coeffB})--(\ref{coeffmu}).

It is worth noting, however, that we are not allowed to deform the Kerr metric in an arbitrary
way. We recall indeed that the Hamiltonian 
that we have derived in Sec.~\ref{sec:hamiltonian_kerr} 
is only valid for a stationary axisymmetric metric, and in coordinates which are related to quasi-isotropic 
coordinates by a redefinition of the radius. In other words, it must be
possible for our deformed metric to be put in the form~\eqref{eq:metric} by a coordinate change of the type $R=R(r)$. 
For this reason we cannot deform the metric exactly 
in the same way as in Ref.~\cite{DJS08}. Here we propose to deform the metric potentials 
in the following manner
\begin{eqnarray}
\label{eq:B}
B&=&\frac{\sqrt{\Delta_t}}{R}\,,\\
\omega&=&\frac{\widetilde{\omega}_{\rm fd}}{\Lambda_t}\,,\label{eq:omega}\\
e^{2\nu}&=&\frac{\Delta_t\,\Sigma}{\Lambda_t}\label{eq:nu}\,,\\
e^{2\mu}&=&\frac{\Sigma}{R^2}\label{eq:mu}\,,
\end{eqnarray}
and
\begin{equation}
J^{-1}= \frac{d R }{d r} = \frac{R}{\sqrt{\Delta_r}}\,,\label{eq:newJ}
\end{equation}
where the relation between $r$ and $R$ can be found by integrating Eq.~\eqref{eq:newJ}:
\begin{equation}
\label{eq:integralR}
R=\exp\left({\int \frac{\mbox{d}r}{\sqrt{\Delta_r}}}\right)\,.
\end{equation}
The deformed metric therefore takes the form
\begin{subequations}
\begin{eqnarray}
\label{def_metric_in}
g^{tt} &=& -\frac{\Lambda_t}{\Delta_t\,\Sigma}\,,\\
g^{rr} &=& \frac{\Delta_r}{\Sigma}\,,\\
g^{\theta\theta} &=& \frac{1}{\Sigma}\,,\\
g^{\phi\phi} &=& \frac{1}{\Lambda_t}
\left(-\frac{\widetilde{\omega}_{\rm fd}^2}{\Delta_t\,\Sigma}+\frac{\Sigma}{\sin^2\theta}\right)\,,\label{eq:gff}\\
g^{t\phi}&=&-\frac{\widetilde{\omega}_{\rm fd}}{\Delta_t\,\Sigma}\,,\label{def_metric_fin}
\end{eqnarray}
\end{subequations}
which does not depend on $R$. Therefore, as we will show explicitly later in this section, 
we do \textit{not} need to compute
the integral~\eqref{eq:integralR} to write the Hamiltonian.
The quantities $\Delta_t$, $\Delta_r$, $\Lambda_t$ and $\widetilde{\omega}_{\rm fd}$ 
in Eqs.~(\ref{def_metric_in})--(\ref{def_metric_fin}) are given by 
\begin{eqnarray}
\label{deltat}
\Delta_t &=& r^2\, \left [A(u) + \frac{a^2}{M^2}\,u^2 \right ]\,, \\
\label{deltar}
\Delta_r &=& \Delta_t\, D^{-1}(u)\,,\\
\Lambda_t &=& \varpi^4 - a^2\,\Delta_t\,\sin^2\theta \,,\\
\widetilde{\omega}_{\rm fd}&=& 2 a\, M\, r
+ \omega_1^{\rm fd}\,\eta\,
\frac{a M^3}{r}
+ \omega_2^{\rm fd}\,\eta\,
\frac{M a^3}{r}
\label{eq:omegaTilde}\,,
\end{eqnarray}
where $u = M/r$, $\omega_1^{\rm fd}$ and $\omega_2^{\rm fd}$ are adjustable parameters which regulate the {\it strength} of 
the frame-dragging, and through 3PN order~\cite{Buonanno00,DJS3PN}
\begin{align}
\label{A_PN}
& A(u) = 1 - 2 u + 2 \eta\, u^3 + \eta\,\left (\frac{94}{3} - \frac{41}{32} \pi^2\right)\, u^4\,,
\\
& D^{-1}(u) = 1 + 6 \eta\, u^2 + 2 (26 - 3 \eta)\, \eta\, u^3\,.\label{D_PN}
\end{align}
We find that our deformed metric is the same as the deformed metric of
Ref.~\cite{DJS08}, except for $g^{\phi\phi}$ and $g^{t\phi}$.\footnote{Ref.~\cite{DJS08} 
chooses $g^{\phi\phi}=(-a^2\,\sin^2\theta + \Delta_t)/
(\Delta_t\,\Sigma \sin^2\theta)$ and $ g^{t\phi}=a\,(\Delta_t - \varpi^2)/(\Delta_t\,\Sigma)$,
which are different from our expressions \eqref{eq:gff} and \eqref{def_metric_fin} even for $\omega_1^{\rm fd}=\omega_2^{\rm fd}=0$.}
As we prove below, the differences between our deformation and the deformation 
of Ref.~\cite{DJS08} appear in the Hamiltonian at PN orders higher than 3PN. 

To obtain the total Hamiltonian (\ref{Htotal}), that is $H = H_{\rm NS} + H_{\rm S}$, 
we first compute the Hamiltonian $H_{\rm NS}$ for a non-spinning particle in the deformed-Kerr metric. 
Using Eq.~(\ref{HNSKerr}) and Ref.~\cite{DJS3PN}, we have
\begin{equation}
{H}_{\rm NS} = \beta^i \, p_i + \alpha \sqrt{m^2 + \gamma^{ij}\,p_i\,p_j + {\cal Q}_4(p)}\,,
\label{eq:Hnsdef}
\end{equation} 
where ${\cal Q}_4(p)$ is a term which is quartic in the space momenta $p_i$ and which was introduced in Ref.~\cite{DJS3PN}, and 
\begin{eqnarray}
\label{alpha}
\alpha &=& \frac{1}{\sqrt{-g^{tt}}}\,,\\
\beta^i &=& \frac{g^{ti}}{g^{tt}}\,,\\
\gamma^{ij} &=& g^{ij}-\frac{g^{ti}\,g^{tj}}{g^{tt}}\,.
\label{gamma}
\end{eqnarray}
In Eqs.~(\ref{alpha})--(\ref{gamma}) the metric components have to be replaced 
with those of the deformed-Kerr metric (\ref{def_metric_in})--(\ref{def_metric_fin}).
When expanded in PN orders, Eq.~(\ref{eq:Hnsdef}) coincides, through 3PN order,  
with the Hamiltonian of a non-spinning test particle in the deformed-Kerr metric 
given by Ref.~\cite{DJS08}. 
 
Second,  to calculate $H_{\rm S}$ given by Eqs.~\eqref{Ht}, \eqref{Ht1} and \eqref{Ht2}, 
we need to compute the derivatives of the metric potentials. We obtain
\begin{subequations}
\begin{eqnarray}
\label{eq:Br}
B_r &=& \frac{\sqrt{{\Delta_r}}\,{\Delta_t}^\prime-2 {\Delta_t}}{2 \sqrt{{\Delta_r}\,{\Delta_t}}\,R}\,,\\
\omega_r
&=& \frac{-\Lambda_t^\prime\,\widetilde{\omega}_{\rm fd} + \Lambda_t\,
\widetilde{\omega}^\prime_{\rm fd}}{\Lambda_t^2}\label{eq:omegar}\,,\\
\nu_r &=& \frac{r}{{\Sigma}}+\frac{{\varpi^2}\,\left({\varpi^2}\,{\Delta_t}^\prime-4 r\, 
{\Delta_t}\right)}{2 {\Lambda_t}\, {\Delta_t}}\label{eq:nur}\,,\\
\mu_r &=&\frac{r}{{\Sigma}}-\frac{1}{\sqrt{{\Delta_r}}}\label{eq:mur}\,,\\
B_{\cos\theta}&=&0\,,\\
\omega_{\cos\theta}
&=&-\frac{2 a^2\, \cos\theta\, \Delta_t\, \widetilde{\omega}_{\rm fd}}{\Lambda_t^2}\label{eq:omegacos}\,,\\
\nu_{\cos\theta}&=&\frac{a^2\, {\varpi^2}\, \cos \theta ({\varpi^2}-{\Delta_t})}{{\Lambda_t}\,
{\Sigma}}\label{eq:nucos}\,,\\
\mu_{\cos\theta}&=&\frac{a^2\, \cos\theta}{\Sigma}\label{eq:mucos}\,,
\end{eqnarray}
\end{subequations}
where the prime denotes derivatives with respect to $r$. 
As already stressed, although the metric potentials $B$, $\omega$, $\nu$ and $\mu$ depend on $R$, the 
factors $R$ cancel out in the deformed-Kerr metric. Therefore, those factors must cancel out also 
in $H_{\rm S}$. This happens because 
the reference tetrad field $\tilde{e}_{A}$ which, together with the metric, completely
determines the Hamiltonian [see Eq.~\eqref{Hamiltonian_canonical_final2}], 
 can be defined independently of $R$. 
Indeed, this turns out to be the case, and if we introduce the rescaled potentials
\begin{eqnarray}\label{tildeBr}
\tilde{B} &=& B\, R= \sqrt{\Delta_t}\,,\\
\tilde{B}_r &=& B_r\, R=\frac{\sqrt{{\Delta_r}}\,{\Delta_t}'-2 {\Delta_t}}
{2 \sqrt{{\Delta_r}\,{\Delta_t}}}\,,\\
e^{2\tilde{\mu}} &=& e^{2\mu}\, R^2 = \Sigma\,,\label{tildeMU}\\
\tilde{J} &=&J\, R=\sqrt{\Delta_r}\label{tildeJ}
\end{eqnarray}
and define 
\begin{equation}\label{Qpert}
Q=1+\frac{\Delta_r (\boldsymbol{\hat{p}}\cdot \boldsymbol{n})^2}{\Sigma}+
\frac{(\boldsymbol{\hat{p}}\cdot \boldsymbol{\xi}\, r)^2 \Sigma }{\Lambda_t\,\sin^2\theta}+
\frac{(\boldsymbol{\hat{p}}\cdot \boldsymbol{v}\, r)^2 }{\Sigma\,\sin^2\theta}\,,
\end{equation}
the Hamiltonian $H_{\rm S}$ for the deformed-Kerr metric takes exactly the
same form as in the Kerr case [see Eqs.~\eqref{KerrHS}, \eqref{KerrHSO} and \eqref{KerrHSS}], 
where we recall that $\xi^2=\sin^2\theta$ and where  now $\omega$ and its derivatives, $\nu$ and its derivatives, and
the derivatives of $\mu$ are given by Eqs.~\eqref{eq:omega}, 
\eqref{eq:nu}, and Eqs.~(\ref{eq:Br})--(\ref{eq:mucos}). 
Also, as we have already stressed, in order to express the Hamiltonian $H_{\rm S}$
in a cartesian coordinate system
in which the spin of the deformed-Kerr black hole is not directed along the $z$-axis, it is  
sufficient to replace $r$ with $(x^2+y^2+z^2)^{1/2}$, 
$\cos\theta$ with $\boldsymbol{\hat{S}}_{\rm Kerr}\cdot \boldsymbol{n}$, 
$\boldsymbol{e}_z$ with $\boldsymbol{\hat{S}}_{\rm Kerr}$ in Eq.~\eqref{eq:xidef},
and to express the vectors appearing in the Hamiltonian in terms of their cartesian
components.

\subsection{PN  expansion of the deformed Hamiltonian}
\label{sec:pn}

We now expand the deformed Hamiltonian $H=H_{\rm NS}+H_{\rm S}$ derived in the previous section
into PN orders. We will denote the spin of the deformed-Kerr metric with 
$\boldsymbol{S}_{\rm Kerr}$, while for the test particle's spin
we introduce the rescaled spin vector $\boldsymbol{S}^\ast=\boldsymbol{S}\, M/m$,
$\boldsymbol{S}$ being the physical, unrescaled spin.
Also, we rescale the spins as $\boldsymbol{S}_{\rm Kerr} \rightarrow \boldsymbol{S}_{\rm Kerr}\, c$ and  
$\boldsymbol{S}^\ast \rightarrow \boldsymbol{S}^\ast\, c$, so as 
to keep track of the PN orders correctly. Moreover, we set 
$\boldsymbol{S}_{\rm Kerr}=\boldsymbol{\chi}_{\rm Kerr}\, M^2$, $\boldsymbol{\chi}_{\rm Kerr}$ being 
the dimensionless spin of the deformed-Kerr black hole, with 
norm $|\boldsymbol{\chi}_{\rm Kerr}|$ ranging from 0 to 1.

As already mentioned, the part of the Hamiltonian which 
does not depend on the test particle's spin, 
$H_{\rm NS}$, agrees through 3PN order with the corresponding $H_{\rm NS}$ computed 
in Ref.~\cite{DJS08}. Moreover, although the metric 
\eqref{def_metric_in}--\eqref{def_metric_fin} only coincides with the
Kerr metric for $\eta=0$, the dependence on $\eta$ appears neither in
the 2PN order coupling of the deformed-Kerr black hole's spin with itself, nor in
its 1.5PN and 2.5PN order spin-orbit couplings. Those couplings are therefore the 
same as in the case of the Kerr metric, and they are given by
\begin{align}
\label{eq:HNSSO}
&H^{\rm NS}_{\rm SO\,1.5PN} = \; \frac{1}{c^3}\,\frac{2}{r^3} \, \boldsymbol{L}\cdot \boldsymbol{S}_{\rm Kerr} \,,\\
\label{eq:HNSSObis}
&H^{\rm NS}_{\rm SO\,2.5PN} = \; 0 \,,\\
\label{eq:HNSSS}
&{H}^{\rm NS}_{\rm SS\,2PN} = \; \frac{1}{c^4}\,\frac{m}{2M\,r^3}\,(3n_{i}\, n_j 
- \delta_{ij})\,S^i_{\rm Kerr}\,S^j_{\rm Kerr}\nonumber\\&\qquad\qquad\qquad\nonumber -\frac{1}{c^4}\,\frac{m}{2 M r^{3}}\,\left[(\boldsymbol{n} \cdot \boldsymbol{S}_{\rm Kerr})^2 - 
 \boldsymbol{S}_{\rm Kerr}^2\right]\\ &\qquad\qquad\qquad+ \frac{1}{2 m (M r)^2 c^4}\Big\{-[\boldsymbol{p}^2 -
 2 (\boldsymbol{p}\cdot \boldsymbol{n})^2] \boldsymbol{S}_{\rm Kerr}^2 \nonumber \\&\qquad\qquad\qquad+
[(\boldsymbol{p}-2 (\boldsymbol{p}\cdot \boldsymbol{n})\boldsymbol{n})\cdot\boldsymbol{S}_{\rm Kerr}]\, 
\boldsymbol{p}\cdot\boldsymbol{S}_{\rm Kerr}\Big\}\,.
\end{align}
Expanding then in PN orders the part of the Hamiltonian that depends on the  test particle's spin, 
that is $H_{\rm S}$, we find
\begin{eqnarray}
  {H}^{\rm S}_{\rm SO\,1.5PN} &=& \frac{3}{2r^3\,c^3} \, \boldsymbol{L}\cdot {\boldsymbol{S}}^\ast\,,
  \label{eq:H15PN}\\
{H}^{\rm S}_{\rm SO\,2.5PN} &=& \frac{1}{r^3\,c^5}\, \left[-\frac{M}{r}
    \left(\frac12 + 3\eta\right) - \frac{5}{8} \boldsymbol{\hat{p}}^2
  \right]\,
  \boldsymbol{L}\cdot\boldsymbol{S}^\ast \nonumber\\
  &&\!\!\!\!\!\!\!\! +\frac{[\boldsymbol{S}^\ast\cdot (\boldsymbol{S}_{\rm
      Kerr}\times \boldsymbol{n})]\,[\boldsymbol{S}_{\rm
      Kerr}\cdot(\boldsymbol{p}-2 (\boldsymbol{p}\cdot
    \boldsymbol{n})\, \boldsymbol{n})]}{M^3\,
    r^3\,c^5}\nonumber\\
&& \!\!\!\!\!\!\!\! +\frac{(\boldsymbol{L}\cdot\boldsymbol{S}^\ast)
    \boldsymbol{S}^2_{\rm Kerr} -(\boldsymbol{L}\cdot\boldsymbol{S}_{\rm
      Kerr})\,
    (\boldsymbol{S}^\ast\cdot\boldsymbol{S}_{\rm
      Kerr})}{2 M^3\, r^4\,c^5}\,.\nonumber \\\label{H25PNnew}\\
  {H}^{\rm S}_{\rm SS\, 2PN} &=& \frac{m}{M\,r^3\,c^4}(3n_{i}\, n_j -
  \delta_{ij})\,S^i_{\rm Kerr}\,
  S_\ast^j \,.\label{eq:H_SS} 
\end{eqnarray}
We recall that the Hamiltonian for a spinning test particle in curved
spacetime from which we started the derivation of our novel EOB model
[see Eq.~\eqref{Hamiltonian_canonical_final2}] is only valid at linear
order in the particle's spin. Therefore, the same restriction applies
to the Hamiltonian derived in Sec.~\ref{sec:kerrdeformed}. In
particular, that Hamiltonian does not include the couplings of the
particle's spin with itself.  We introduce those couplings by hand, 
at least at the leading order (2PN), by adding a quadrupole deformation~\cite{Damour01c} 
$h^{\mu\nu}$, quadratic in the particle's spin, to the 
deformed-Kerr metric in Sec.~\ref{sec:kerrdeformed} [see 
Eqs.~\eqref{def_metric_in}--\eqref{def_metric_fin}].  The expression
for $h^{\mu\nu}$ and the details of the above procedure --- together
with a way in which it can in principle be extended to reproduce also
the next-to-leading order coupling of the particle's spin with itself
--- are given in Appendix~\ref{sec:ss}. For the purpose of the present
discussion, however, it is sufficient to mention that the addition of
this quadrupole deformation to the metric \eqref{def_metric_in}--\eqref{def_metric_fin} augments
Eq.~(\ref{eq:HNSSS}) by the term
\begin{equation}
  \frac{m}{2M\,r^3\,c^4}(3n_{i}\, n_j -
  \delta_{ij})\,S^i_{\ast}\,
  S_\ast^j \,.\label{eq:H_SSstar}
\end{equation}
Therefore, the total leading order spin-spin Hamiltonian is
\begin{align}
H_{\rm SS\,2PN}&={H}^{\rm S}_{\rm SS\, 2PN}+{H}^{\rm NS}_{\rm SS\, 2PN}\nonumber\\&  +\frac{m}{2 M\,r^3\,c^4}(3n_{i}\, n_j -
  \delta_{ij})\,S^i_{\ast}\, S_\ast^j\nonumber\\&=\frac{m}{2M\,r^3\,c^4}(3n_{i}\, n_j -
  \delta_{ij})S_0^iS_0^j\nonumber\\&\nonumber-\frac{1}{c^4}\,\frac{m}{2 M r^{3}}\,\left[(\boldsymbol{n} \cdot \boldsymbol{S}_{\rm Kerr})^2 - 
 \boldsymbol{S}_{\rm Kerr}^2\right]\\ &+ \frac{1}{2 m\,M^2\, r^2\, c^4}\Big\{-[\boldsymbol{p}^2 -
 2 (\boldsymbol{p}\cdot \boldsymbol{n})^2] \boldsymbol{S}_{\rm Kerr}^2 \nonumber \\&+
[(\boldsymbol{p}-2 (\boldsymbol{p}\cdot \boldsymbol{n})\boldsymbol{n})\cdot\boldsymbol{S}_{\rm Kerr}]\, 
\boldsymbol{p}\cdot\boldsymbol{S}_{\rm Kerr}\Big\}\label{eq:H_SS_total}
\,,
\end{align}
with $S_0^i=S^i_{\rm Kerr}+S^i_{\ast}$.

As we will show in Sec.~\ref{sec:EOB}, a proper choice of the vectors
$\boldsymbol{S}_{\rm Kerr}$ and $\boldsymbol{S}^\ast$ in terms of the vectors 
$\boldsymbol{\sigma}$ and $\boldsymbol{\sigma}^\ast$, defined in
Eqs.~(\ref{sigma}) and (\ref{sigmastar}), allows
us to reproduce the PN-expanded effective Hamiltonian [see Eqs.~(\ref{eq:HeffSO})--(\ref{eq:HeffSS})] 
using the PN-expanded deformed-Kerr Hamiltonian that we have just derived. 

Finally, it is worth noting that the presence of terms quadratic in the 
deformed-Kerr black hole's spin in Eq.~(\ref{H25PNnew}) explains why we introduced 
the 2.5PN-accurate canonical transformation~(\ref{G25PNnew}). Indeed, the latter 
produces exactly the same terms in the PN-expanded effective Hamiltonian~(\ref{eq:HeffSO}) 
at 2.5PN order. Quite interestingly, the terms quadratic in $\boldsymbol{S}_{\rm Kerr}$ 
appearing in Eq.~(\ref{H25PNnew})
could also be eliminated with a suitable choice of the reference tetrad
$\tilde{e}_A$. In fact, as stressed in Sec.~\ref{sec:Hamiltonian_axisymmetric} 
and in Ref.~\cite{diracbrackets_ours}, a choice of the reference tetrad field
corresponds to choosing a particular gauge for the particle's spin.  
In agreement with this interpretation, we find that the terms of Eq.~(\ref{H25PNnew}) 
which are quadratic in $\boldsymbol{S}_{\rm Kerr}$
disappear if the initial tetrad \eqref{tetrad0}--\eqref{tetrad3} is
changed to a different tetrad $\tilde{\boldsymbol{e}}^{\prime}_{A}$
related to the original one by the following purely-spatial rotation:
\begin{equation}
\tilde{\boldsymbol{e}}^{\prime T} 
= \tilde{\boldsymbol{e}}^{T} \,,\quad \quad  
\tilde{\boldsymbol{e}}^\prime_I = {\cal R}_{IJ}\, 
\tilde{\boldsymbol{e}}_J 
\label{tetrad}\,,
\end{equation}
where the rotation matrix ${\cal R}_{IJ}$ is given by
\begin{equation}
{\cal R}={\cal R}_Y\left [-\frac{a^2\, X\, Z}{2 R^4} \right ]\, 
{\cal R}_X \left [-\frac{a^2\, Y\, Z}{2 R^4} \right ]\,,
\end{equation}
${\cal R}_X[\psi]$ and ${\cal R}_Y[\phi]$ being 
rotations of angles $\psi$ and $\phi$ around the axis $X$ and $Y$, respectively. 

As a consistency check, we have verified that this new tetrad is the
same as that used in Ref.~\cite{diracbrackets_ours} when computing the
Hamiltonian in ADM coordinates, where those terms quadratic in
$\boldsymbol{S}_{\rm Kerr}$ do not appear.  We have checked this by
transforming the new tetrad~\eqref{tetrad} from quasi-isotropic to ADM
coordinates [which are related by the coordinate transformation~(49)
in Ref.~\cite{Hergt:2007ha}], and comparing it to the tetrad given in
Eqs.~(6.9a)--(6.9b) of Ref.~\cite{diracbrackets_ours}, and find that
the two tetrads agree through order $1/c^8$.

\subsection{The effective-one-body Hamiltonian}
\label{sec:EOB}

In this section we first find the mapping between the masses 
$\mu$, $M$ and the spins $\boldsymbol{\sigma}$ and $\boldsymbol{\sigma}^\ast$ 
of the effective Hamiltonian derived in Sec~\ref{sec:eff}, 
and those of the deformed-Kerr Hamiltonian derived in Secs.~\ref{sec:kerrdeformed}
and~\ref{sec:pn}, that is $m$, $M$, $\boldsymbol{S}_{\rm Kerr}$ and 
$\boldsymbol{S}^\ast$. Then, we derive the improved (resummed) EOB Hamiltonian. 

As shown in Ref.~\cite{Buonanno00}, matching the non-spinning parts
$H_{\rm NS}$ of these Hamiltonians forces us to identify the total
mass $M$ of the two black holes in the PN description  
with the deformed-Kerr mass $M$ of the test-particle description, 
thus justifying our choice of using the same
symbol for these two \textit{a priori} distinct quantities. Similarly, we
find that $m=\mu$~\cite{Buonanno00}.  Assuming this 
mapping between the masses and imposing that the PN-expanded 
deformed-Kerr Hamiltonian given by 
Eqs.~(\ref{eq:HNSSO})--(\ref{eq:H_SS_total}) coincides with the
effective Hamiltonian given by Eqs.~(\ref{eq:HeffSO})--(\ref{eq:HeffSS}), 
we obtain the following mapping between the spins
\begin{eqnarray}
\label{mapping1}
\boldsymbol{{S}}^\ast &=& \boldsymbol{\sigma}^\ast+\frac{1}{c^2}\,\boldsymbol{\Delta}_{{\sigma}^\ast}\,,\\
\label{mapping2}
\boldsymbol{{S}}_{\rm Kerr} &=& \boldsymbol{\sigma}+\frac{1}{c^2}\,\boldsymbol{\Delta}_{{\sigma}}\,,
\end{eqnarray}
where we have set for simplicity $a(\eta)=0$ and $b(\eta)=0$ and where
\begin{eqnarray}
\boldsymbol{\Delta}_{{\sigma}}&=&-\frac{1}{16}\,\Bigg\{
12 \boldsymbol{\Delta}_{{\sigma}^\ast}
+\eta \bigg [ \frac{2 M}{r}\, (4 \boldsymbol{\sigma}
-7 \boldsymbol{\sigma}^\ast)  \nonumber \\
&&  + 6 (\hat{\boldsymbol{p}}\cdot\boldsymbol{n})^2 
\,(6 \boldsymbol{\sigma}+5 \boldsymbol{\sigma}^\ast)
-\boldsymbol{\hat{p}}^2\,(3 \boldsymbol{\sigma}+4 \boldsymbol{\sigma}^\ast)
\bigg]\Bigg\}\,.
\nonumber \\\label{deltasigma}
\end{eqnarray}
Here, $\boldsymbol{\Delta}_{{\sigma}^\ast}$ is an arbitrary function
going to zero at least linearly in $\eta$ when $\eta\to0$, so as to
get the correct test-particle limit.  In fact, if
$\boldsymbol{\Delta}_{{\sigma}^\ast}$ satisfies this condition and if
we assume, as appropriate for black holes,
$\boldsymbol{S}_{1,2}=\boldsymbol{\chi}_{1,2}\, m_{1,2}^2$ (with $\vert
\boldsymbol{\chi}_{1,2} \vert\leq1$ and constant)\footnote{As noted by
  Ref.~\cite{DJS08}, a spin mapping such as ours also gives the
  correct test particle limit if $\vert
  \boldsymbol{S}_{1,2}\vert/m_{1,2}= $ const., but this scaling of the
  spins with the masses is not appropriate for black
  holes~\cite{hartl}.}, when $m_2\sim 0$ we have $\boldsymbol{S}_{\rm
  Kerr}=\boldsymbol{S}_{1}+{\cal O}(m_2)$.  
Similarly, for $m_2\sim 0$ the physical unrescaled spin of the effective particle
is $\boldsymbol{S}=\boldsymbol{S}^\ast\, m/M=\boldsymbol{S}_{2}
+{\cal O}(m_2)^2$. The equations of motion
of our initial Hamiltonian~\eqref{Hamiltonian_canonical_final2}
coincide with the  Papapetrou
equations~\cite{diracbrackets_ours}, which describe the motion of a spinning test-particle in a
curved spacetime~\cite{Papa51spin,CPapa51spin}. Assuming the canonical commutation
relations between $x^i$, $p_j$, $\boldsymbol{S}_{1}$ and $\boldsymbol{S}_{2}$, we obtain that 
the Hamilton equations for the effective deformed-Kerr Hamiltonian are 
$\dot{y}=\dot{y}_{\rm P}+{\cal O}(m_2)$.
Here, the dot denotes a time derivative, $y$ is a generic phase-space variable 
($x^i$, $p_j$, $\boldsymbol{S}_{1}$ or $\boldsymbol{S}_{2}$),
and $\dot{y}=\dot{y}_{\rm P}$ are the Papapetrou equations expressed in Hamiltonian form. 
Therefore, our mapping reproduces
the correct test-particle limit, and the remainders  $\boldsymbol{S}_{\rm Kerr}-\boldsymbol{S}_{1}={\cal O}(m_2)$
and $\boldsymbol{S}-\boldsymbol{S}_{2}={\cal O}(m_2)^2$ produce extra-accelerations of order ${\cal O}(m_2)$ or higher.
This is comparable to the self-force acceleration~\cite{poisson_self_force_review}, which appears at the
next order in the mass ratio beyond the test-particle limit.

Although different choices for the function $\boldsymbol{\Delta}_{{\sigma}^\ast}$ 
are in principle possible, we choose here
\begin{eqnarray}
\boldsymbol{\Delta}_{{\sigma}^\ast}&=&
\frac{\eta}{12}\,\Big [ \frac{2M}{r}\,(7 \boldsymbol{\sigma}^\ast-4 
\boldsymbol{\sigma})+ \boldsymbol{\hat{p}}^2 \,
(3 \boldsymbol{\sigma}+4 \boldsymbol{\sigma}^\ast)  \nonumber\\ 
&&  -6 (\boldsymbol{\hat{p}}\cdot\boldsymbol{n})^2\,(6 \boldsymbol{\sigma}+5 
\boldsymbol{\sigma}^\ast)\Big ]\,,
\end{eqnarray}
which gives, when inserted into Eq.~\eqref{deltasigma},
$\boldsymbol{\Delta}_{\sigma}=0$. Because this form for
$\boldsymbol{\Delta}_{{\sigma}^\ast}$ is clearly not covariant under
generic coordinate transformations, we choose
instead the following form for the mapping of the spins, which is
covariant at least as far as the square of the momentum is concerned:
\begin{align}
\boldsymbol{\Delta}_{\sigma}=&\;0\,,\\
\boldsymbol{\Delta}_{\sigma^\ast}=&\;\frac{\eta}{12}\,  
\Big [ \frac{2M}{r}\,(7 \boldsymbol{\sigma}^\ast-4 \boldsymbol{\sigma})+ 
(Q-1)\, (3 \boldsymbol{\sigma}+4 \boldsymbol{\sigma}^\ast)\nonumber\\ &
-6 \frac{\Delta_r}{\Sigma}\,
(\hat{\boldsymbol{p}}\cdot\boldsymbol{n})^2\, (6 \boldsymbol{\sigma}+5 \boldsymbol{\sigma}^\ast)\Big]\,,
\end{align}
where we have replaced $\hat{\boldsymbol{p}}^2$ with
$\gamma^{ij}\hat{p}_i\hat{p}_j=Q-1$ [where $Q$ is given in
Eq.~\eqref{Qpert}] and
$(\hat{\boldsymbol{p}}\cdot\boldsymbol{n})^2=\hat{p}_r^2$ with
$\Delta_r(\hat{\boldsymbol{p}}\cdot\boldsymbol{n})^2/\Sigma=g^{rr}\hat{p}_r^2$.
This form agrees with the previous mapping through order $1/c^2$, but
differs from it at higher orders. Although neither this form is
completely covariant, not even under a rescaling of the radial
coordinate (as it still features a dependence on the radius $r$), it
proved slightly better as far as the dynamics of the EOB model,
analyzed in the next section, is concerned. In particular, the factor
$g^{rr}$, which becomes zero at the horizon, quenches the increase of
$\hat{p}_r$ at small radii, thus giving a more stable behavior during
the plunge subsequent to the inspiral. (A similar effect was observed
in Ref.~\cite{Pan2009}, where the radial momentum was expressed in
tortoise coordinates to prevent it from diverging close to the
horizon.)

Having determined the mass and spin mappings, we can write down the 
improved (resummed) Hamiltonian (or EOB Hamiltonian) for spinning black holes. To this
purpose, it is sufficient to invert the mapping between the real and
effective Hamiltonians [Eq.~\eqref{heff}]. In units in which $c=1$, we
obtain
\begin{equation}
\label{hreal}
H_\mathrm{real}^{\rm improved} = M\,\sqrt{1+2\eta\,\left(\frac{H_{\rm eff}}{\mu}-1\right)}\,,
\end{equation}
with 
\begin{eqnarray}
\label{HeffEOB}
H_{\rm eff} &=& {H}_{\rm S}+ \beta^i\,p_i+ \alpha\, \sqrt{\mu^2 + \gamma^{ij}\,p_i\,p_j + {\cal Q}_4(p)} \nonumber \\
&& -\frac{\mu}{2M\, r^3}\,(\delta^{ij} - 3 n^i\,n^j)\, S^\ast_i\, S^\ast_j\,.
\nonumber \\
\end{eqnarray}
Here, the $-{\mu}/{(2M\, r^3)}(\delta^{ij} - 3 n^in^j) S^\ast_i S^\ast_j$ term is the 
quadrupole deformation introduced in the previous section
to account for the leading order coupling of the particle's spin with itself (see also Appendix~\ref{sec:ss});
$\beta^i$, $\alpha$ and $\gamma^{ij}$ are computed using the deformed-Kerr 
metric, that is inserting Eqs.~(\ref{def_metric_in})--(\ref{def_metric_fin}) 
into Eqs.~(\ref{alpha})--(\ref{gamma}); $H_{\rm S}$ 
is obtained by inserting Eqs.~\eqref{eq:omega}, \eqref{eq:nu}, and Eqs.~(\ref{eq:Br})--(\ref{Qpert})
into Eqs.~\eqref{KerrHS}, \eqref{KerrHSO} and \eqref{KerrHSS}. 
Lastly, the spin $\boldsymbol{S}_{\rm Kerr}$ enters this Hamiltonian through the 
parameter $a=|\boldsymbol{S}_{\rm Kerr}|/M$ appearing in the deformed-Kerr metric.

Before completing this section, we want to discuss the deformation of the 
Kerr potentials $\Delta_t$ and $\Delta_r$ given in Eqs.~(\ref{deltat}) and 
(\ref{deltar}), which play an important role in the EOB Hamiltonian (\ref{hreal}).  
It is convenient to re-write the function $\Delta_t$ as  
\begin{eqnarray}
\label{deltatu}
\Delta_t &=& r^2\, \Delta_u(u)\,, \\
\Delta_u(u) &=& A(u) + \frac{a^2}{M^2} u^2\,.
\label{deltauu}
\end{eqnarray}
In previous EOB investigations the Pad\'e summation was applied 
to the function $\Delta_u$ to enforce the presence of a zero, 
corresponding to the EOB horizon, both 
in the non-spinning~\cite{DJS3PN} and 
spinning case~\cite{Damour01c,DJS08}. Reference~\cite{Pan2009} pointed out 
that when including the 4PN and 5PN terms in the function $A(u)$, the 
Pad\'e summation generates poles if spins are present. Also,   
the Pad\'e summation does not always ensure the existence of an innermost stable circular orbit (ISCO) for 
spins aligned and antialigned with the orbital angular momentum and, even when it does, 
the position of the ISCO does not vary monotonically with the magnitude of 
the spins. For these reasons, we propose here an alternative way of enforcing the existence of 
the EOB horizons. Working through 3PN order, we write 
\begin{eqnarray}
\label{delta_t_1}
\Delta_u(u) &=& \bar{\Delta}_u(u)\, \left [1 + \eta\,\Delta_0 + 
\eta \,\log \left (1 + \Delta_1 \,u + \Delta_2\,u^2 \right. \right . 
\nonumber \\
&& \left. \left. + \Delta_3\,u^3 + \Delta_4\,u^4\right ) \right ]\,,
\end{eqnarray}
where 
\begin{align}
\bar{\Delta}_u(u)=&\,\frac{a^2}{M^2}\,\left(u - \frac{M}{r^{\rm EOB}_{\rm H,+}}\right)\,
\left(u - \frac{M}{r^{\rm EOB}_{\rm H,-}}\right) \\=&\,\frac{a^2 u^2}{M^2}+\frac{2 u}{\eta  K-1}+\frac{1}{(\eta  K-1)^2}\,,\\
\label{eq:hor}
r^{\rm EOB}_{\rm H,\pm} =&\, \left(M \pm \sqrt{M^2 - a^2}\right)\, (1 - K\,\eta)\,.
\end{align}
Here, $r^{\rm EOB}_{\rm H,\pm}$ are the EOB horizons, which differ from the Kerr horizons 
when the adjustable parameter $K$ is different from zero,
and where the $\log$ is introduced to quench the divergence of the powers of
$u$ at small radii. We could in principle replace the logarithm with any other 
analytical function with no zeros (e.g., an exponential).
However, when studying the dynamics of the EOB model (see Sec.~\ref{sec:hamiltonian_eob}) the results 
are more sensible
if we choose a function, such as the logarithm, which softens the divergence of the truncated PN series.

The coefficients $\Delta_0$, $\Delta_1$, $\Delta_2$, $\Delta_3$ and $\Delta_4$  can be derived by inserting 
Eq.~(\ref{delta_t_1}) into Eq.~(\ref{deltatu}), expanding through 3PN order, and equating the 
result to Eqs.~(\ref{deltatu}) and (\ref{deltauu}), with $A(u)$ given
by its PN expansion~(\ref{A_PN}). Doing so, we obtain
\begin{widetext}
\begin{eqnarray}
\Delta_0 &=& K\,(\eta\,K - 2)\label{eq:k_0}\,,\\
\Delta_1 &=& -2(\eta\,K - 1)\,(K+\Delta_0)\,,\\
\Delta_2 &=& \frac{1}{2}\,\Delta_1\, (-4 \eta\,K +\Delta_1 +4)-\frac{a^2}{M^2}\,(\eta\,K-1)^2\, \Delta_0\,,\\
\Delta_3 &=& \frac{1}{3}\,\Big [-\Delta_1^3+ 3 (\eta \,K-1)\, \Delta_1^2+3 \Delta_2\, \Delta_1-6 (\eta \,K-1) \,
(-\eta \,K + \Delta_2+1) -3 \frac{a^2}{M^2}\, (\eta\,K-1)^2\, \Delta_1\Big ]\,,\\
\Delta_4 &=& \frac{1}{12}\Big\{6 \frac{a^2}{M^2}\, \left(\Delta_1^2-2 \Delta_2\right) (\eta\,K-1)^2+ 3 \Delta_1^4-
8(\eta\,K-1)\,\Delta_1^3-12 \Delta_2\,\Delta_1^2+12 \left [2 (\eta\,K-1)\,\Delta_2+\Delta_3\right]\, \Delta_1  \nonumber\\
&&  +12 \left (\frac{94}{3} - \frac{41}{32} \pi^2\right)\,(\eta\,K-1)^2+6 \left [\Delta_2^2-4 \Delta_3 \,(\eta\,K -1)\right ]\Big\}\,.
\label{eq:k_4}
\end{eqnarray}
\end{widetext}
By construction, if we expand Eq.~(\ref{delta_t_1}) in PN orders, $K$
can only appear at 4PN and higher orders, because we must recover
the PN expansion~(\ref{deltat})--\eqref{A_PN} through 3PN order. 
In this sense, $K$ parameterizes our ignorance of 
the PN expansion at orders equal or higher than 4PN
(i.e., $K$ would not play any role if the PN series were
known in its entirety).
Similarly, we re-write the potential $\Delta_r$ [Eq.~(\ref{deltar})] as 
\begin{eqnarray}
\label{eq:D}
\Delta_r &=& \Delta_t\,D^{-1}(u)\,,\label{eq:deltaR}\\
D^{-1}(u) &=& 1+\log[1 + 6 \eta\, u^2 + 2 (26 - 3 \eta)\, \eta\, u^3]\,.\nonumber \\
\end{eqnarray}
The coefficients in the above function $D^{-1}(u)$ are such that, 
when PN expanded, it gives the PN result~\eqref{D_PN}, and the 
logarithmic dependence is once again
chosen to quench the divergence of the truncated PN series.

Finally, let us stress that if we included PN orders higher 
than 3PN in the functions $A(u)$ and $D(u)$, we would need to add higher order 
coefficients $\Delta_i$ with $i > 4$ in Eq.~(\ref{delta_t_1}).

\section{Effective-one-body dynamics for circular,  equatorial  orbits}
\label{sec:hamiltonian_eob}

In this section we study the dynamics of the novel EOB 
model that we developed in Sec.~\ref{sec:EOB}. We will show that 
\begin{enumerate}
\item[(i)] Our EOB model has the correct test-particle limit, for both non-spinning and spinning 
black holes, for \textit{generic} orbits and \textit{arbitrary} spin orientations; 
\item[(ii)] There exist an ISCO when the spins are aligned or antialigned with the orbital 
angular momentum $\boldsymbol{L}$;
\item[(iii)] The radius, energy, 
total angular momentum, orbital angular momentum and frequency at the ISCO 
exhibit a smooth dependence  on the binary mass-ratio and spins. Also, this 
dependence looks reasonable based on what we expect from the test-particle limit and 
from numerical-relativity simulations;
\item[(iv)] The frequency at the ISCO for an extreme mass-ratio non-spinning 
black-hole binary agrees with the exact result computed by Ref.~\cite{barack_sago};
\item[(v)] During the plunge subsequent to the ISCO, the orbital frequency of black-hole binaries 
with spins aligned or antialigned with $\boldsymbol{L}$ grows and reaches a maximum, 
after which it decreases. The radius at which the frequency peaks is very close to the 
radius of the equatorial, circular light ring (or photon orbit). 
This feature generalizes the non-spinning behavior~\cite{Buonanno00}, and it 
has a clear physical interpretation in terms of frame-dragging. As in the 
non-spinning case~\cite{Buonanno00}, it provides a natural 
time at which to match the two-body description of 
the inspiral and plunge to the one-body description of the merger and ringdown.
\end{enumerate}
We stress that only {(i)} applies to generic orbits and spin orientations, while {(ii)}, {(iii)}, {(iv)}
and {(v)} are true for black-hole binaries with spins aligned or antialigned 
with $\boldsymbol{L}$. (It should be noted that circular or spherical
orbits, and therefore the ISCO, are not even present for generic
orbits and spin orientations, because the system is not 
integrable, not even in the test-particle limit~\cite{hartl}). 
While we will tackle the study of generic orbits and arbitrary spin orientations in a follow-up
paper, we argue that the preliminary study
presented here is already sufficient to illustrate the potential of
the novel EOB model. We recall~\cite{Pan2009} that the only
existing EOB model for spinning black-hole binaries, proposed in 
Refs.~\cite{Damour01c,DJS08}, {(i)} reproduces only approximately the test
particle limit; {(ii)} when including non-spinning terms at 4PN and 
5PN order, it does not always present an ISCO for
binaries with spins parallel to $\boldsymbol{L}$, and when it does 
the spin dependence of quantities evaluated at the ISCO is unusual; 
{(iii)} generally, the orbital frequency does not peak during the plunge, 
making the prediction of the matching time from the two-body to the one-body 
description quite problematic.  

Let us now go through the points of the list that we presented at the beginning of this section. 
In order to prove point {(i)} we first need to observe
that the deformed  metric~\eqref{def_metric_in}--\eqref{def_metric_fin} [with the potentials $\Delta_r$ and 
$\Delta_t$ given by Eqs.~\eqref{eq:deltaR} and~\eqref{delta_t_1}] 
reduces to the Kerr metric as $\eta\to0$, and the deformation is linear in $\eta$ when
$\eta\sim0$. Therefore, the acceleration produced by this deformation on the test-particle is
comparable to the self-force acceleration, which appears at the
next order in the mass ratio beyond the test-particle limit. Second, as already proved in Sec.~\ref{sec:EOB}, 
the mapping~\eqref{mapping1}--\eqref{mapping2} of the spins reduces to 
$\boldsymbol{S}_{\rm Kerr}=\boldsymbol{S}_{1} +{\cal O}(m_2)$ 
and $\boldsymbol{S}=\boldsymbol{S}^\ast m/M=\boldsymbol{S}_{2} +{\cal O}(m_2)^2$ when $m_2\sim0$,
where the remainders produce accelerations which are again comparable to the self-force acceleration.

To prove points {(ii)}, {(iii)}, {(iv)} and  {(iv)}, we need to write the effective EOB 
Hamiltonian (\ref{HeffEOB}) for equatorial orbits and for spins parallel 
to the orbital angular momentum (chosen to be along the $z$-axis). We obtain
\begin{eqnarray}
\label{HeffEOBco}
H_{\rm eff} &=& H_{\rm S} + \beta^i\,p_i+ \alpha \sqrt{\mu^2 + 
\gamma^{ij}\,p_i\,p_j +{\cal Q}_4(p)}
 \nonumber \\
&& -\frac{\mu}{2M\, r^3}\,\, S_\ast^2
\,,\nonumber \\\\
 H_{\rm S}&=&g^{\rm eff}_{\rm SO}\,\boldsymbol{L}\cdot\boldsymbol{S}^\ast +g^{\rm eff}_{SS}\, {S}^\ast\,,\nonumber \\
\end{eqnarray}
where
\begin{widetext}
\begin{eqnarray}
g^{\rm eff}_{\rm SO} &=&\frac{e^{2 \nu -\tilde{\mu}}\, \left[-\sqrt{Q\,\Delta_r}\, (\tilde{B}_r-2 \tilde{B}\,\nu_{r})
+\left(e^{\tilde{\mu}+\nu }-\tilde{B}\right)\, \left(\sqrt{Q}+1\right)+(\tilde{B}\,\nu_{r}-\tilde{B}_r)
 \sqrt{\Delta_r}\right]}{\tilde{B}^2 \,M \left(\sqrt{Q}+1\right)\, \sqrt{Q}}\,,\\
 g^{\rm eff}_{\rm SS}&=&\frac{\mu}{M}\,\left[\omega+\frac{1}{2}\, \tilde{B}\, e^{-\tilde{\mu}-\nu }\, \omega_{r}\,\sqrt{\Delta_r}
+\left(\frac{L_z^2}{\mu^2}-\tilde{B}^2\, e^{-2 (\tilde{\mu}+\nu )}\, \Delta_r\, \frac{p^2_r}{\mu^2}\right)
\frac{e^{\nu -\tilde{\mu}}\, \omega_{r}\, \sqrt{\Delta_r}}{2 \tilde{B}\, \left(\sqrt{Q}+1\right) \,\sqrt{Q}}\right]\,,
\end{eqnarray}
\end{widetext}
with
\begin{equation}\label{Qeob}
Q=1+\frac{\Delta_r\, p_r^2}{\mu^2\, r^2}+\frac{L_z^2\, r^2}{\mu^2\, (\varpi^4 - a^2\,\Delta_t)}\,.
\end{equation}
The above equations can be evaluated explicitly by using Eqs.~\eqref{eq:omega}, \eqref{eq:nu},
(\ref{eq:Br})--(\ref{eq:mur}), 
(\ref{tildeBr})--(\ref{tildeMU}), (\ref{deltatu})-- (\ref{eq:D})\footnote{A Mathematica notebook implementing the Hamiltonian \eqref{HeffEOBco}--\eqref{Qeob} is available from the authors upon request.}. 
To calculate the radius and the orbital angular momentum at the ISCO for the 
EOB model, we insert Eq.~(\ref{HeffEOBco}) into the real EOB Hamiltonian (\ref{hreal}), 
and solve numerically the following system of equations~\cite{Buonanno00}
\begin{eqnarray}
&& \frac{\partial H^{\rm improved}_\mathrm{real}(r,p_r=0,L_z)}{\partial r}=0\,, \\
&& \frac{\partial^2 H^{\rm improved}_\mathrm{real}(r,p_r=0,L_z)}{\partial r^2}=0\,,
\end{eqnarray}
with respect to $r$ and $L_z$. Moreover, the frequency for 
circular orbits is given by
\begin{equation}
\Omega=\frac{\partial H^{\rm improved}_\mathrm{real}(r,p_r=0,L_z)}{\partial L_z}\,,
\end{equation}
which follows immediately from the Hamilton equations because $L_z=p_\phi$. Finally  
the binding energy is $E_{\rm bind}=H^{\rm improved}_\mathrm{real}-M\,$.

Henceforth, we set the adjustable frame-dragging parameters 
$\omega^{\rm fd}_1=\omega^{\rm fd}_2=0$ [see Eq.~\eqref{eq:omegaTilde}]
and write $K$ in Eq.~\eqref{eq:hor} as a polynomial of second order in $\eta$, 
\begin{equation}
K(\eta) = K_0+K_1\, \eta + K_2\, \eta^2\,.
\end{equation}
$K_0$, $K_1$ and $K_2$ being constants.
We find that if we impose
\begin{equation}
K(1/4) = \frac{1}{2}\,, \quad \quad  \frac{d K}{d \eta}(1/4)=0 
\end{equation} 
the functional dependence on $\eta$ and $\chi$ 
of several physical quantities evaluated at the ISCO is quite 
smooth and regular. Therefore, imposing these constraints we obtain
\begin{equation}
\label{simpleK} 
K(\eta) =K_0\, (1-4 \eta)^2+ 4 (1-2 \eta) \eta\,.  
\end{equation}
It is worth noting that the values of $K$ and $d K/d \eta$ at
$\eta=1/4$ have a more direct meaning than the coefficients $K_1$ and
$K_2$.  In fact, current numerical-relativity simulations can evolve
binary black holes with $\eta \approx 0.25$ (with only few runs having
$\eta\sim 0.1$). Thus, they can determine $K(1/4)-1/2$, while the
value of $(dK/d\eta)(1/4)$ can be hopefully determined when more
numerical simulations with $\eta = 0.1 \mbox{--} 0.25$ become available.

Hereafter, we will use Eq.~\eqref{simpleK} and set $K_0=1.4467$. The latter is 
determined by requiring that the ISCO frequency 
for extreme mass-ratio non-spinning black-hole binaries agrees with the exact result 
of Ref.~\cite{barack_sago}, which computed the shift of the ISCO frequency due to the conservative part of 
self-force (see also Ref.~\cite{damour_SF} where the result of Ref.~\cite{barack_sago} 
was compared to the non-spinning EOB prediction which resums the function (\ref{A_PN}) 
{\it \`a la} Pad\'e.).\footnote{We stress that the most general form of $K(\eta)$ can include 
terms depending on $a^2$, with $a=|\boldsymbol{S}_{\rm Kerr}|/M$. 
In particular, a term \textit{not} depending on $\eta$ and proportional to $a^2$ could be
  determined by a calculation similar to that in Ref.~\cite{barack_sago}, that is 
by computing the shift of the ISCO frequency caused by the conservative part  of the self-force, for a non-spinning test-particle 
in a Kerr spacetime.}
\begin{figure}
       \resizebox{80mm}{!}{\includegraphics{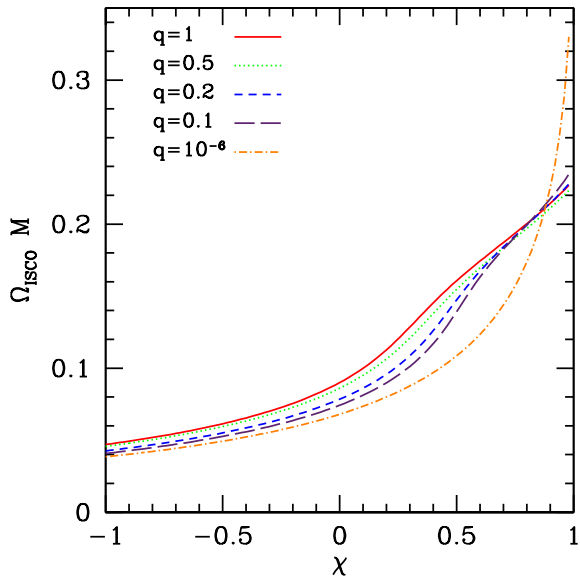}} 
       \caption{The frequency at the EOB ISCO for binaries having spins parallel to $\boldsymbol{L}$, with mass ratio
         $q=m_2/m_1$ and with spin-parameter projections onto the direction of $\boldsymbol{L}$ given by 
	 $\chi_1=\chi_2=\chi$. As expected, the
         frequency increases with $\chi$ for a given mass ratio, while
         for fixed $\chi$ it increases with $q$ if $\chi\lesssim 0.9$,
         while it decreases with $q$ if $\chi$ is almost extremal (see
         text for details).}
\label{freq_isco1}
 \end{figure}
 \begin{figure}
       \resizebox{80mm}{!}{\includegraphics{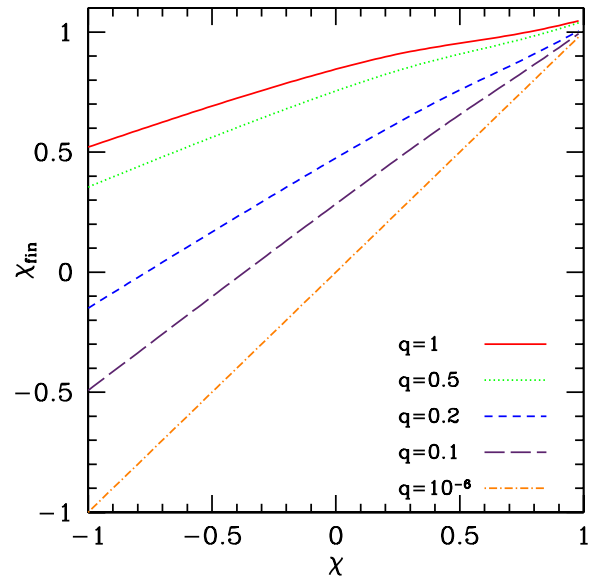}} 
       \caption{The final spin parameter $\chi_{\rm fin}$ as inferred
         at the EOB ISCO, for binaries having spins parallel to $\boldsymbol{L}$, with mass ratio
         $q=m_2/m_1$ and with spin-parameter projections onto the direction of $\boldsymbol{L}$ given by 
	 $\chi_1=\chi_2=\chi$. As expected, $\chi_{\rm fin}$
         flattens for large $\chi$ in the comparable mass case (see text
         for details).}
\label{fig:afin1}
 \end{figure}
 \begin{figure}
       \resizebox{80mm}{!}{\includegraphics{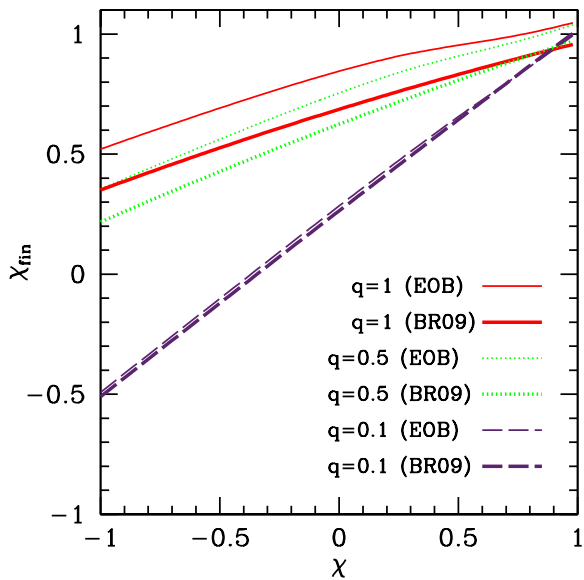}} 
       \caption{The final spin parameter $\chi_{\rm fin}$ as inferred
         at the EOB ISCO for binaries having spins parallel to $\boldsymbol{L}$, with mass ratio
         $q=m_2/m_1$ and with spin-parameter projections onto the direction of $\boldsymbol{L}$ given by 
	 $\chi_1=\chi_2=\chi$, compared to the remnant's final spin
         parameter predicted by the formula presented in
         Ref.~\cite{spin_formula} (``BR09''), which accurately
         reproduces numerical-relativity results. The EOB model and the BR09 
         formula agree when the mass
         ratio is small ($q=0.1$), because the emission during the
         plunge, merger and ringdown is negligible in this case.
	 For $q=0.5$ and $q=1$, there is an offset, 
         because the EOB result, at this stage, neglects the gravitational-wave emission
         during the plunge, merger and ringdown (see text for details).}
\label{fig:afin_vs_aei}
 \end{figure}
\begin{figure}
       \resizebox{80mm}{!}{\includegraphics{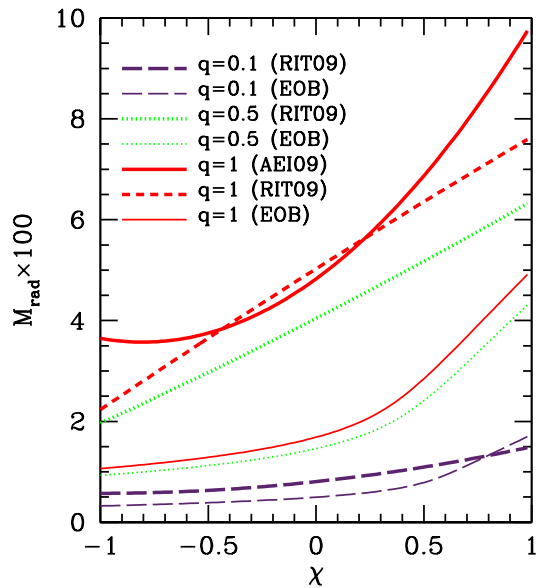}} 
       \caption{The mass loss inferred at the EOB ISCO for binaries having spins parallel to $\boldsymbol{L}$, with mass ratio
         $q=m_2/m_1$ and with spin-parameter projections onto the direction of $\boldsymbol{L}$ given by 
	 $\chi_1=\chi_2=\chi$, 
         compared to the total mass lost during the inspiral, merger
         and ringdown, as predicted by the formulas presented in
         Ref.~\cite{aei_aligned_spins} (``AEI09'') and in
         Ref.~\cite{RITfit} (``RIT09''), which reproduce numerical-relativity results,
         although with different accuracies because of the different
         parameter regions they cover (see the text for
         details). The EOB model and the AEI09 and RIT09 fits agree when the mass
         ratio is small ($q=0.1$), while there is an offset for
         $q=0.5$ and $q=1$. The reason is that the ringdown
         emission, which is negligible for small mass-ratios, is not
         taken into account by our EOB model at this stage.}
\label{fig:Eb_vs_aei}
 \end{figure}
 \begin{figure}
       \resizebox{80mm}{!}{\includegraphics{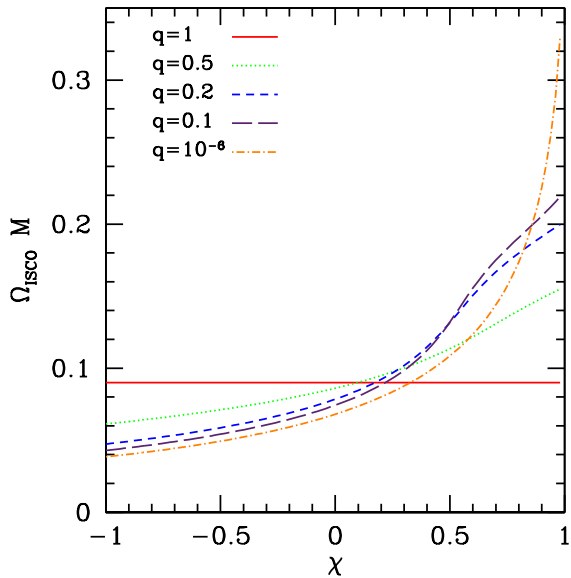}} 
       \caption{The frequency at the EOB ISCO for binaries having spins parallel to $\boldsymbol{L}$, with mass ratio
         $q=m_2/m_1$ and with spin-parameter projections onto the direction of $\boldsymbol{L}$ given by 
	 $\chi_1=-\chi_2=\chi$. As expected, the
         frequency is constant in the equal-mass case, because the
         spins of the two black holes cancel out (see text for details).}
\label{freq_isco2}
 \end{figure}

 \begin{figure}
       \resizebox{80mm}{!}{\includegraphics{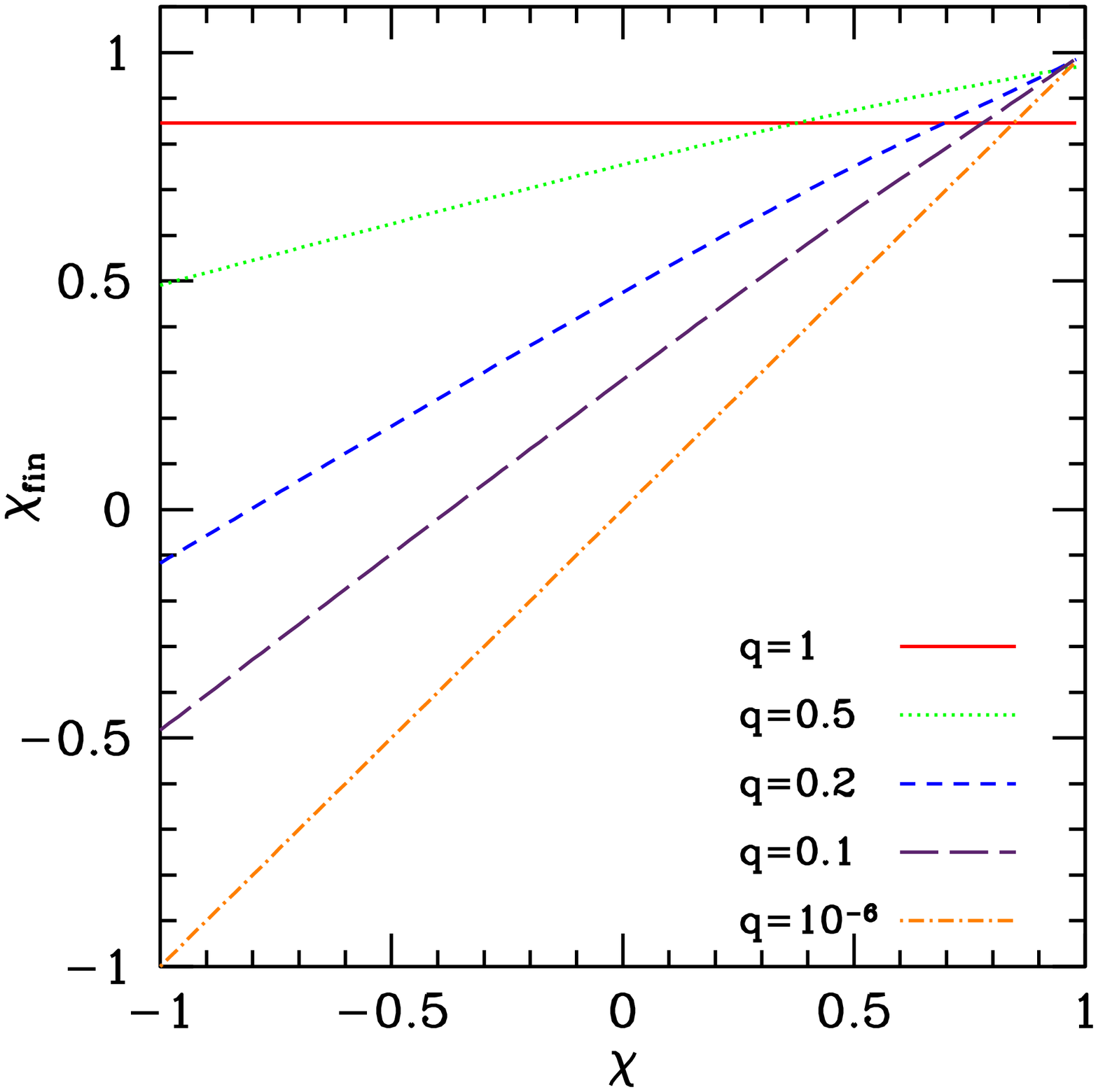}} 
       \caption{The final spin parameter $\chi_{\rm fin}$ as inferred
         at the EOB ISCO, for binaries having spins parallel to $\boldsymbol{L}$, with mass ratio
         $q=m_2/m_1$ and with spin-parameter projections onto the direction of $\boldsymbol{L}$ given by 
	 $\chi_1=-\chi_2=\chi$. The results are the same
         for all equal-mass binaries, for which the spins of
         the two black holes cancel out (see text for details).}
\label{fig:afin2}
 \end{figure}
In Fig.~\ref{freq_isco1} we plot the orbital frequency at the ISCO for binaries 
with mass ratio $q=m_2/m_1$ ranging from $10^{-6}$ to $1$ and spins aligned with
$\boldsymbol{L}$. In particular, denoting by $S_{1,2}=\chi_{1,2} m^2_{1,2}$ the
projections of the spins along the direction of $\boldsymbol{L}$, we
consider binaries with $\chi_1=\chi_2=\chi$. We see that 
the ISCO frequency increases with the magnitude of the spins $\chi$
if the mass ratio is fixed, as expected from the test-particle 
case. Also, if the spins are kept fixed and small, the ISCO frequency
increases with the mass-ratio, as it should be to 
reproduce the results of numerical-relativity simulations
(see, e.g., the non-spinning EOB models of Ref.~\cite{Damour2009a,Buonanno:2009qa}). 
However, if the spins are close to $\chi=1$, the ISCO frequency decreases when the mass
ratio increases. This crossover is mirrored by a similar
behavior of other quantities evaluated at the ISCO --- such as the
 energy, the orbital angular momentum, and the coordinate radius --- and
 its physical meaning can be explained as follows. When 
 comparable-mass almost-extremal black holes merge, the resulting black-hole remnant
 has a spin parameter that is slightly smaller than the spin
 parameters of the parent black holes. This is a consequence of the cosmic censorship
 conjecture (see Ref.~\cite{naked} and references therein) which prevents black holes with spin $\chi>1$ to be formed~\cite{dain}.  
 Therefore, because in the EOB model the position, and therefore the frequency, 
 of the ISCO (together with the loss of energy and angular momentum during the 
 plunge) regulate the final spin of the remnant,
and because for an isolated black hole the ISCO frequency
increases with the spin, any
 EOB model that satisfies the  cosmic censorship
 conjecture must have an ISCO
 frequency that slightly decreases with the mass ratio when
 $\chi \sim 1$.
 \begin{figure}
       \resizebox{80mm}{!}{\includegraphics{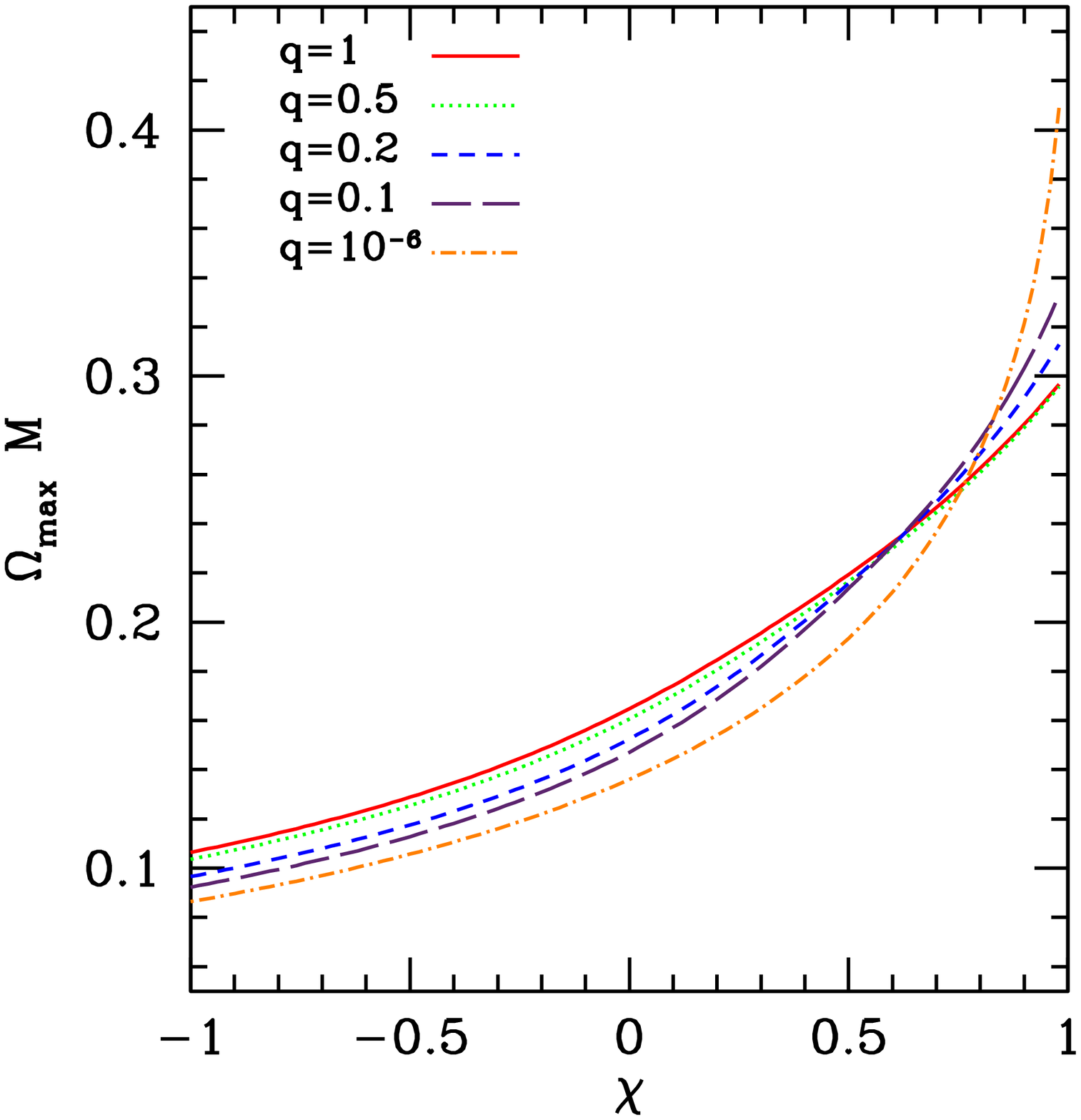}} 
       \caption{The maximum of the EOB orbital frequency during the plunge, 
	 for binaries having spins parallel to $\boldsymbol{L}$, with mass ratio
         $q=m_2/m_1$ and with spin-parameter projections onto the direction of $\boldsymbol{L}$ given by 
	 $\chi_1=\chi_2=\chi$. As expected, the frequency increases
         with $\chi$ for a given mass ratio, while for fixed $\chi$ it
         increases with $q$ if $\chi\lesssim 0.9$, while it decreases
         with $q$ is $\chi$ is almost extremal (see text for
         details).}
\label{fig:freqMax1}
 \end{figure}
 \begin{figure*}
   \begin{center}
     \begin{tabular}{cc}
       \resizebox{80mm}{!}{\includegraphics{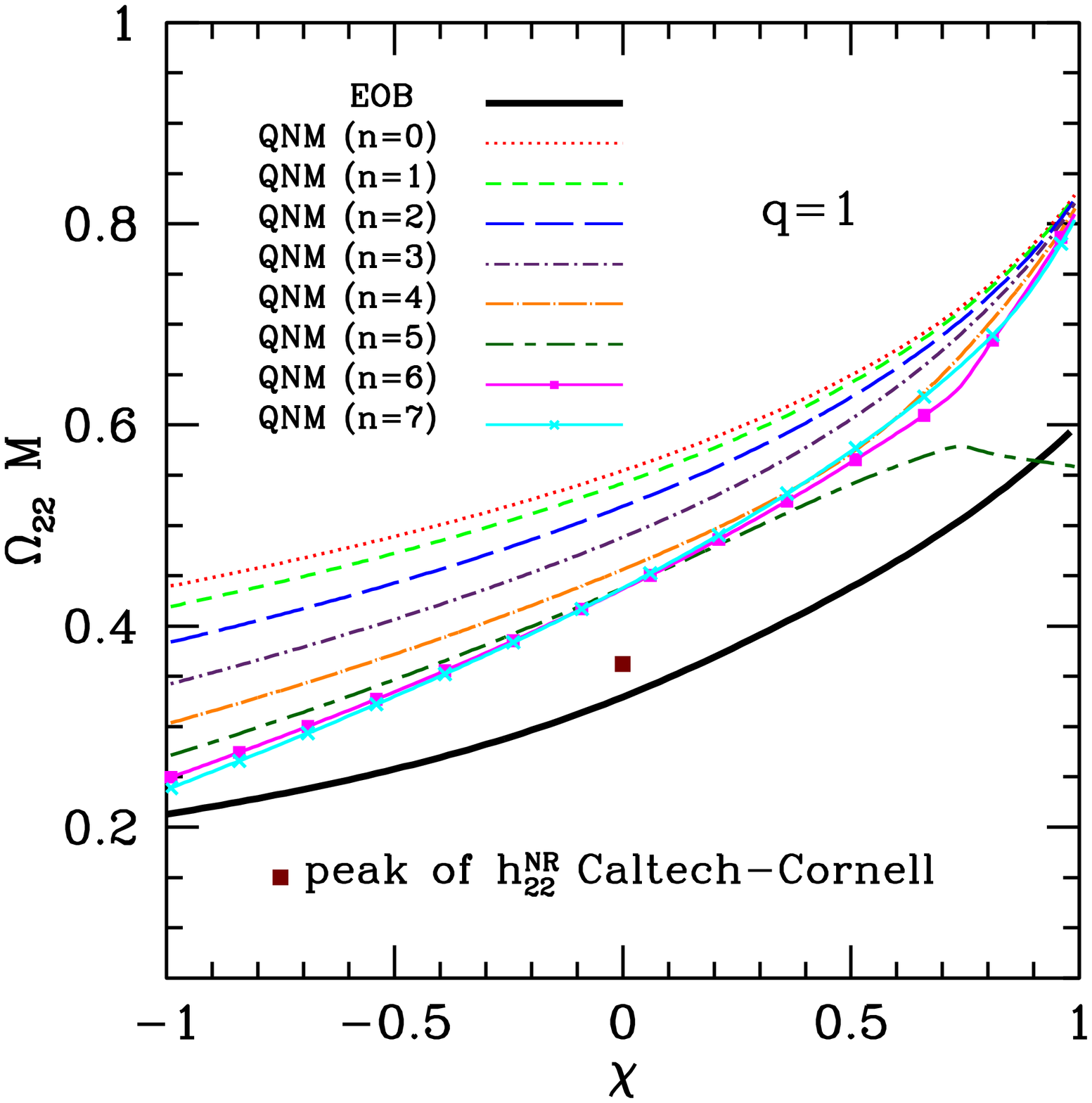}} &
       \resizebox{80mm}{!}{\includegraphics{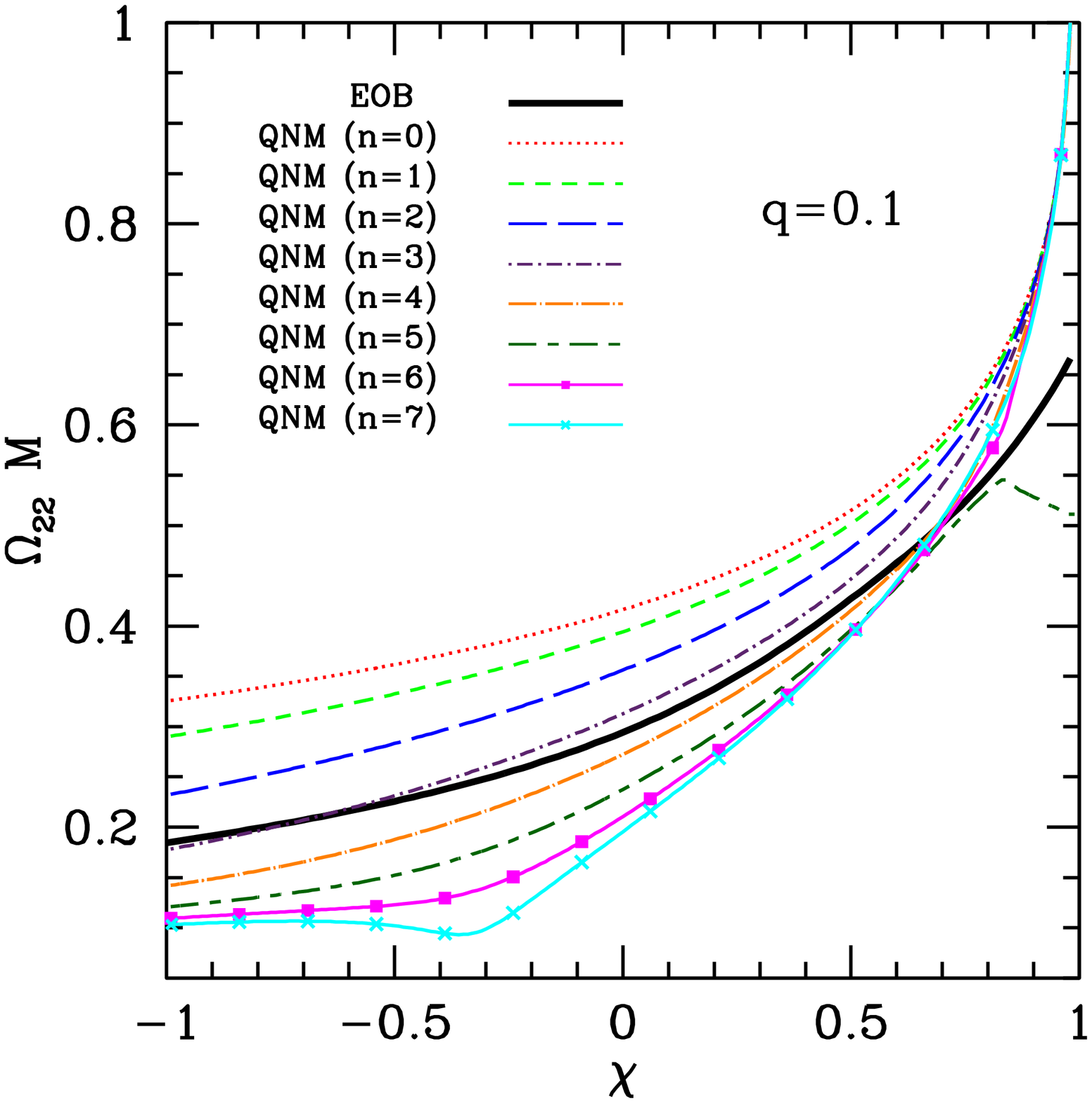}} \\
     \end{tabular}
     \caption{For binaries having spins parallel to $\boldsymbol{L}$, with mass ratio $q=m_2/m_1=1$ (left panel) and
       $q=m_2/m_1=0.1$ (right panel) and with spin-parameter projections onto the direction of $\boldsymbol{L}$ given by 
       $\chi_1=\chi_2=\chi$, we plot twice the maximum of the EOB orbital frequency
       during the plunge against the frequencies of the first 8
       overtones of the $\ell=2$, $m=2$ quasi-normal mode of a Kerr black hole.
       The quasi-normal mode frequency is computed using the final spin and 
       final mass of the remnant. The
       final spin is estimated by applying the formula of
       Ref.~\cite{spin_formula}, while for the final mass we use the
       formula of Ref.~\cite{aei_aligned_spins} (``AEI09'') in the
       $q=1$ case and that of Ref.~\cite{RITfit} (``RIT09'') in the
       $q=0.1$ case. 
       As can be seen, in the $q=0.1$ case the peak frequencies lie among
       the high overtones of the $\ell=2$, $m=2$ mode, while in the
       $q=1$ case they are generally lower than them. In the $q=1$ case we also mark with a square the numerical 
       gravitational-wave frequency at the peak of the $h_{22}$ mode when $\chi=0$. 
       This gravitational-wave frequency coincides with (twice) the maximum of the EOB orbital frequency 
       at the time when the matching of the quasi-normal modes is performed
       in the non-spinning EOB model of Ref.~\cite{Buonanno:2009qa}. The numerical gravitational-wave 
       frequency is computed from the numerical simulation of  
       Refs.~\cite{caltech_cornell_nonspinning}
       (``Caltech-Cornell'').}
\label{qnm} 
\end{center}
 \end{figure*}
This interpretation can be confirmed by computing the final spin of the remnant black hole as estimated at the ISCO. 
We have
\begin{equation}
\chi_{\rm fin}=\frac{S_1+S_2+L_{{\rm ISCO}}}{(M+E^{\rm bind}_{{\rm ISCO}})^2}\,,
\end{equation}
which is plotted in Fig.~\ref{fig:afin1}. Although the final spin gets
slightly larger than $1$ for high initial spins (because we are
neglecting here the energy and angular momentum emitted during the
plunge, merger and ringdown), the curves are remarkably smooth and monotonic
(see the corresponding Fig. 5 of Ref.~\cite{DJS08})
and they flatten at high initial spins, as expected. In particular, in
Fig.~\ref{fig:afin_vs_aei} we focus on mass ratios $q=1$, $q=0.5$
and $q=0.1$, and plot the final spin $\chi_{\rm fin}$ as
inferred from the ISCO energy and angular momentum, together with the
final spin of the remnant predicted by the formula presented in
Ref.~\cite{spin_formula} which accurately reproduce the numerical-relativity results
(see also Refs.~\cite{aei1,aei2,aei3,aei4, BKL,fau,RITfit} for other formulas for the final spin
of the remnant). It is remarkable that in spite of the offset between
the predictions of the formula of Ref.~\cite{spin_formula} 
and the EOB result, which is due to neglecting
the energy and angular momentum emitted during plunge, merger and ringdown, 
the qualitative behavior of the curves in Fig.~\ref{fig:afin_vs_aei} 
is the same. Also, we observe that the difference between 
corresponding curves decreases with 
the mass ratio, with the EOB and the numerical-relativity--based results 
being in very good agreement for $q=0.1$. This happens because the 
energy and angular momentum emitted during plunge, merger and ringdown 
become negligible for small mass-ratios.\footnote{This can be seen by noting that, for a test-particle 
with mass $m$ around a black holes with mass $M$, the final plunge
  lasts a dynamical time $\sim M$~\cite{Buonanno00}, while the inspiral from 
 large radii to the ISCO lasts $\sim M^2/m$.}. Similarly, in Fig.~\ref{fig:Eb_vs_aei} we 
plot the binding energy at the ISCO for mass ratios $q=1$, $q=0.5$
and $q=0.1$ and compare it with fits to numerical-relativity data 
for the total mass radiated in gravitational waves during the inspiral, merger and
ringdown.  In particular, for the $q=1$ case we use the fit in 
Ref.~\cite{aei_aligned_spins} (``AEI09''), while for
$q=0.5$ and $q=0.1$ we use the fit recently proposed by
Ref.~\cite{RITfit} (``RIT09''). While the AEI fit is more accurate
than the RIT one for the particular configuration considered here
(see Fig. 11 and related discussion in Ref.~\cite{aei_aligned_spins}), the AEI fit
is only applicable for comparable-mass binaries\footnote{The limited
applicability of the AEI fit (which is only valid for equal-mass binaries with spins aligned or anti-aligned)
is indeed one reason why it turns out to be more accurate, for the configuration under 
consideration, than the RIT fit, which is instead applicable to more generic binaries.}, and for this reason
we resort to the more general RIT fit in the $q=0.5$ and
$q=0.1$ cases. [In Fig.~\ref{fig:Eb_vs_aei} we show the predictions of both the AEI and the
RIT fit in the $q=1$ case. Being the AEI fit more accurate, its difference 
from the RIT fit gives an idea of the error bars which should be applied to the predictions
of the RIT fit for $q=0.5$ and $q=0.1$.]  

 \begin{figure}
       \resizebox{80mm}{!}{\includegraphics{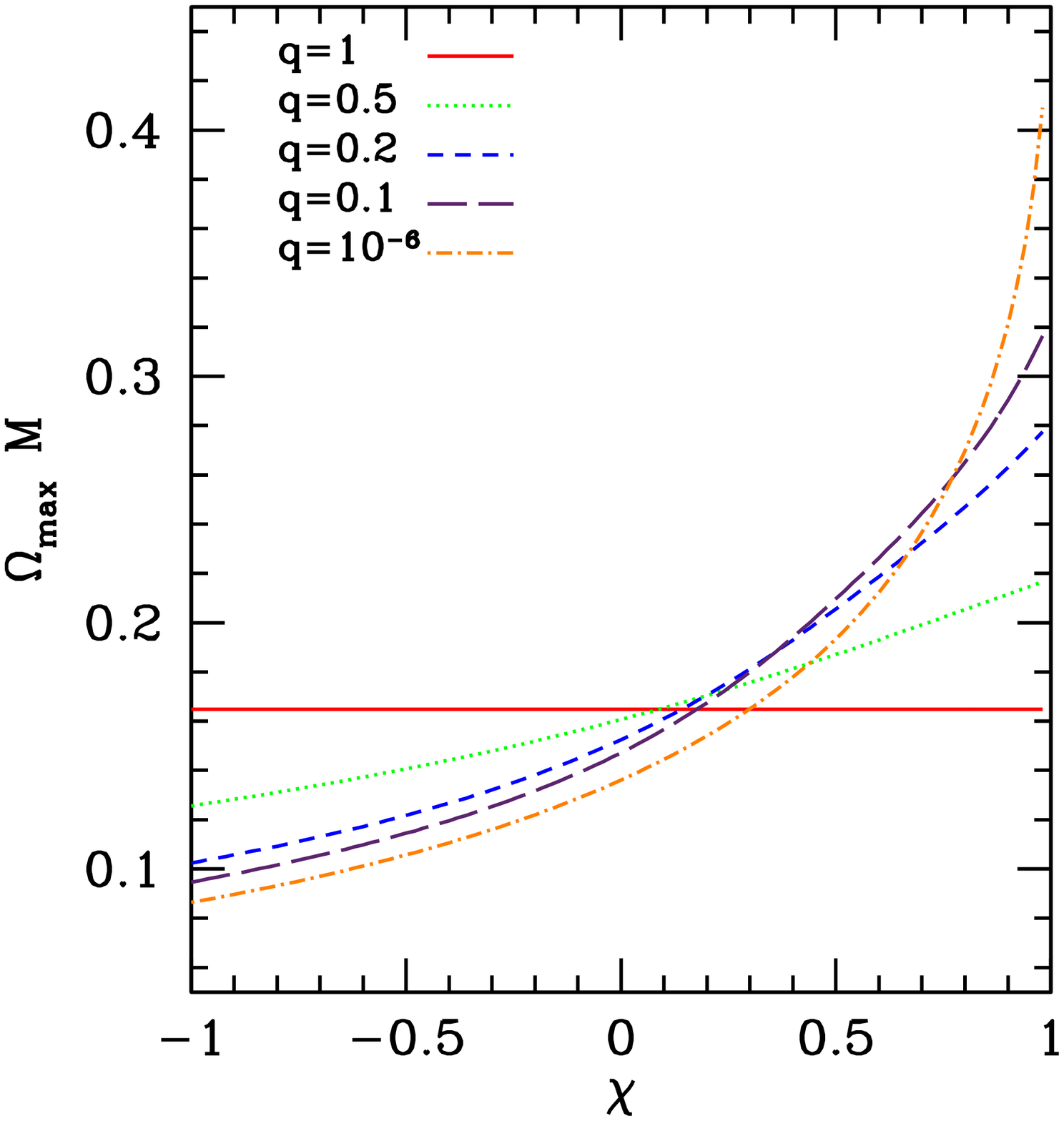}} 
       \caption{The maximum of the EOB orbital frequency during the plunge, for binaries 
	 having spins parallel to $\boldsymbol{L}$, with mass ratio
         $q=m_2/m_1$ and with spin-parameter projections onto the direction of $\boldsymbol{L}$ given by 
	 $\chi_1=-\chi_2=\chi$. The results are the same for all
         equal-mass binaries, for which the spins of the two
         black holes cancel out (see text for details).}
\label{fig:freqMax2}
 \end{figure}
In Figs.~\ref{freq_isco2} and~\ref{fig:afin2} we present similar results, for the ISCO frequency and for the final spin
estimated at the ISCO,
in the case of spins antialigned with the orbital angular momentum.  The
 most apparent feature of these figures is that, in the equal-mass
 case, the quantities under consideration are independent of $\chi$.
 This happens because in this case the spins $S_1$ and $S_2$ are equal and
 opposite, which results in a zero value for the spins $S_{\rm Kerr}$
 and $S^\ast$  entering the EOB Hamiltonian~\eqref{hreal}. As such,
 in the EOB model, equal-mass binaries with equal and opposite spins behave as non-spinning
 binaries.  This feature, which is also shared by the PN-expanded Hamiltonian, 
 until the PN order which is currently known, is also in 
 agreement with the results of numerical simulations. In fact, 
 equal-mass binaries with equal and opposite spins would be 
 indistinguishable with LISA, Virgo and LIGO observations~\cite{aei_aligned_spins,vaishnav}.  
 Except for this feature, and similarly to the aligned case discussed above, 
 the behavior of the curves in Figs.~\ref{freq_isco2}
 and~\ref{fig:afin2} is quite smooth and regular when 
 going from the equal-mass case to the test-particle case. 

 In Fig.~\ref{fig:freqMax1} we plot the maximum value of the
 orbital frequency during the plunge subsequent to the inspiral,
 for binaries with mass ratio $q=m_2/m_1$ and with
 $\chi_1=\chi_2=\chi$.  More precisely, we assume that the particle
 starts off with no radial velocity at the ISCO (thus having angular
 momentum $L_{\rm {ISCO}}$ and energy $E_{\rm {ISCO}}$), and we compute 
 $p_r$ assuming that the energy and angular momentum are conserved during the 
plunge. We find that the orbital frequency presents a peak for any value of the spins and
 any mass ratio, and we denote the value of the frequency at the peak with $M \Omega_{\max}$. 
We note that the behavior of
 $M \Omega_{\max}$ as a function of the mass ratio is similar to 
 that of $M \Omega_{_{\rm ISCO}}$. In particular, its dependence on
 $\eta$ changes sign when going from small to large spins. 

The  physical interpretation of the peak of the orbital frequency 
is that the frequency increases as the effective particle spirals in, 
but when the effective particle gets close to the black hole, 
the orbital frequency has to decrease because the particle's motion gets 
locked to the horizon (this is a well-known phenomenon, 
see for instance Ref.~\cite{skimming1,skimming2} for
some of its effect on the test-particle dynamics). Said in another 
way, the orbital frequency of the effective particle for an 
observer at infinity goes to a constant (or to zero in 
the non-spinning case~\cite{Buonanno00}) on the EOB horizon.
As a consequence, the peak in the frequency can be used to signal 
the transition between two regimes~\cite{Buonanno00}:
one in which the deformed black hole and the effective particle have different 
frequencies and one in which the two bodies basically move and radiate as a single 
perturbed black hole. For this reason the peak of the frequency provides the 
EOB approach with the natural point where to switch to the one-body description, 
i.e., the point where to start describing the gravitational waveforms as a 
superposition of quasi-normal modes. 

We find that the values of $M \Omega_{\max}$ are roughly those needed to attach the quasi-normal
 modes used in EOB models to describe the merger and the
 ringdown~\cite{Damour2009a,Buonanno:2009qa,Pan2009}. This is shown in
 Fig.~\ref{qnm}, where we plot twice the maximum of the orbital frequency, $M \Omega_{22}$ 
for binaries with $q=1$ (left panel) and $q=0.1$ (right panel) and with spins $\chi_1=\chi_2=\chi$. 
 We compare $M \Omega_{22}$ with the frequency of the first 8 overtones of the $\ell=2$,
 $m=2$ quasi-normal mode of a Kerr black hole, computed using 
  the final spin and the final mass of the black-hole remnant~\cite{berti}. [The final
 spin parameter is estimated by applying the formula of
 Ref.~\cite{spin_formula}, while for the final mass we use the formula
 of Ref.~\cite{aei_aligned_spins} (``AEI09'') in the $q=1$ case and
 the one of Ref.~\cite{RITfit} (``RIT09'') in the $q=0.1$ case.] 
In the $q=1$ case we also mark with a square the numerical 
       gravitational-wave frequency at the peak of the $h_{22}$ mode when $\chi=0$. 
       This gravitational-wave frequency coincides with (twice) the maximum of the EOB orbital frequency 
       at the time when the matching of the quasi-normal modes is performed
       in the non-spinning EOB model of Ref.~\cite{Buonanno:2009qa}. The numerical gravitational-wave 
       frequency is computed from the numerical simulation of  
       Refs.~\cite{caltech_cornell_nonspinning}
       (``Caltech-Cornell''). As can be seen, while in the $q=0.1$ case the
 peak frequencies lie among the high overtones of the $\ell=2$, $m=2$
 quasi-normal mode, in the $q=1$ case they are generally lower than
 them. Quite interestingly, we find that the values of $M \Omega_{22}$ 
for $\chi\gtrsim0.4$ can be increased up to the frequencies of the quasi-normal modes by
 assuming $\omega^{\rm fd}_2 \sim 30\mbox{--}70$ in Eq.~\eqref{eq:omegaTilde}.  Nevertheless,
 the frequencies that we obtain are comparable to those used for the
 matching with the quasi-normal modes in Ref.~\cite{Buonanno:2009qa,Pan2009}, and we
 therefore expect such a matching to be possible also in our EOB
 model.

Also, it is interesting to note that the position $r_{\max}$ of the frequency
peak is quite close (to within $8 \%$) to the position of the light ring (or 
circular photon orbit). This fact, which holds exactly in the non-spinning case~\cite{Buonanno00},
further confirms that the potential barrier for massless particles
(such as gravitational waves) lies at $r \sim  r_{\max}$.

 Finally, in Fig.~\ref{fig:freqMax2} we show the maximum value of
 the orbital frequency during the plunge for binaries with mass
 ratio $q=m_2/m_1$ and with $\chi_1=-\chi_2=\chi$. As for the ISCO 
 quantities, the dependence on the spins and the mass ratios is much simpler than in
 the aligned case, with the black-hole spins cancelling out in the equal
 mass-case and thus giving results which are independent of $\chi$.
 Also, we see a smooth transition from the equal-mass to the extreme
 mass-ratio case, that our model reproduces exactly. Also in this antialigned 
 case, the radius $r_{\max}$ agrees with the light-ring position to within $4\%$.

\section{Conclusions}
\label{sec:conclusions}

In this paper, building on Ref.~\cite{diracbrackets_ours}, 
we computed  the Hamiltonian of a spinning test particle, at linear 
order in the particle's spin, in an axisymmetric stationary metric and in
quasi-isotropic coordinates. Then, by applying a coordinate 
transformation, we derived the Hamiltonian of a spinning test particle 
in Kerr spacetime in Boyer-Lindquist coordinates

We used those results to construct an improved EOB Hamiltonian 
for spinning black holes. To achieve this goal, we followed 
previous studies~\cite{Damour01c,DJS08} and mapped the 
real two-body dynamics into the dynamics of an effective 
particle with mass $\mu$ and  spin $\mathbf{S}_\ast$ moving 
in a deformed-Kerr spacetime with spin $\mathbf{S}_{\rm Kerr}$, 
the symmetric mass-ratio of the binary, $\eta$, acting as the deformation parameter.

To derive the improved EOB Hamiltonian, we proceeded  
as follows. First, we applied a suitable canonical transformation 
to the real ADM Hamiltonian and worked out the PN-expanded 
effective Hamiltonian through the relation 
\begin{equation}\label{heff_conclusions}
\frac{H_{\rm eff}}{\mu} =
\frac{H_{\rm real}^2 - m_1^2 - m_2^2}{2 m_1\, m_2}\,. 
\end{equation}
Then, we found an appropriate 
deformed-Kerr metric such that the corresponding Hamiltonian, when 
expanded in PN orders, coincided with the PN-expanded effective Hamiltonian through 
3PN order in the non-spinning terms, and 2.5PN order in the spinning terms. 
 
The (resummed) improved EOB Hamiltonian is then found by inverting Eq.~\eqref{heff_conclusions}, which gives
\begin{equation}
\label{hrealeff}
H_\mathrm{real}^{\rm improved} = M\,\sqrt{1+2\eta\,\left(\frac{H_{\rm eff}}{\mu}-1\right)}\,,
\end{equation}
with
$H_{\rm eff}$ given by Eq.~(\ref{HeffEOB}), where $\alpha$, $\beta^i$ and $\gamma^{ij}$ are
obtained by inserting Eqs.~(\ref{def_metric_in})--(\ref{def_metric_fin}) 
into Eqs.~(\ref{alpha})--(\ref{gamma}); where $H_{\rm S}$ 
is obtained by inserting Eqs.~\eqref{eq:omega}, \eqref{eq:nu}, and Eqs.~(\ref{eq:Br})--(\ref{Qpert})
into Eqs.~\eqref{KerrHS}, \eqref{KerrHSO} and \eqref{KerrHSS}; 
and where the effective particle's spin $\boldsymbol{S}^{\ast}$ and
the deformed-Kerr spin $\boldsymbol{S}_{\rm Kerr}$ (with $a=|\boldsymbol{S}_{\rm Kerr}|/M$) 
are expressed in terms of the real spins by means of Eqs.~(\ref{mapping1}),
(\ref{mapping2}), \eqref{sigma} and \eqref{sigmastar}.

The crucial EOB metric potential for quasi-circular motion is the 
potential $\Delta_t(r)$ (which reduces in the non-spinning case to the 
radial potential $A(r)$ of Refs.~\cite{Buonanno99,Buonanno00}). 
To guarantee the presence of an inner and 
outer horizons in the EOB metric, we proposed to re-write the 
potential $\Delta_t(r)$ in a suitable way [see Eqs.~(\ref{deltatu}) and 
(\ref{delta_t_1})], introducing the adjustable EOB 
parameter $K(\eta)$ regulating the higher-order, unknown PN terms. 
The reason why we did not re-write the potential $\Delta_t(r)$ 
using the Pad\'e summation~\cite{DJS08} is because Ref.~\cite{Pan2009} 
found that when including non-spinning 
terms at 4PN and 5PN order, the Pad\'e summation produces spurious poles, 
does not always ensure the presence of an ISCO for binaries 
with spins parallel to $\boldsymbol{L}$ and, even when it does, 
the spin dependence of physical quantities evaluated at the ISCO 
is quite unusual.

Restricting the study to circular orbits in the equatorial plane and 
assuming spins aligned or antialigned with the 
orbital angular momentum, we investigated several features 
of our improved EOB Hamiltonian. Using an expression of 
the EOB adjustable parameter $K(\eta)$ which reproduces the 
self-force results in the non-spinning extreme mass-ratio limit
~\cite{barack_sago,damour_SF}, we computed the orbital frequency 
at the EOB ISCO, we estimated the final spin from the EOB ISCO, 
and the maximum orbital frequency during the plunge. We found 
that these predictions are quite smooth and regular under a 
variation of $\eta$ and of the black-hole spins. Quite interestingly, 
the maximum of the orbital frequency during the plunge always 
exists and is close to the light-ring position, as in the non-spinning case~\cite{Buonanno00}. 
For this reason, as in the non-spinning case~\cite{Buonanno00}, the orbital-frequency peak can be used within the EOB
to mark the matching time at which the merger and ringdown start, i.e,
 the time when, in the EOB formalism, 
the gravitational waveforms start being described by a superposition 
of quasi-normal modes. This will be useful in future comparisons 
of the EOB model with numerical-relativity 
simulations. 

The results of Sec.~\ref{sec:hamiltonian_eob} are an example of the performances that
our improved EOB Hamiltonian can achieve. We expect several refinements 
to be possibly needed when comparing our EOB model with accurate numerical-relativity 
simulations of binary black holes. We may, for example, extend our model to reproduce also the 
 next-to-leading order spin-spin couplings, which are known and 
appear at 3PN order~\cite{SHS07,SSH08,SHS08,PR06,PR07,PR08b,PR08a}. Also, we might introduce 
a different mapping between the black-hole spins $\mathbf{S}_1$, $\mathbf{S}_2$
and $\mathbf{S}_\ast$, $\mathbf{S}_{\rm Kerr}$, a different form of 
the adjustable parameter $K(\eta)$, and re-write differently the EOB metric 
potential $\Delta_t(r)$. We could also introduce in it the adjustable parameters $a_5$ and 
$a_6$ at 4PN and 5PN order, respectively.  Moreover, other choices of the 
reference tetrad used to work out the Hamiltonian for a spinning test-particle in an 
axisymmetric stationary spacetime could be in principle used, leading to a different 
(canonically related) EOB Hamiltonian. Lastly, the mapping (\ref{heff_conclusions})-(\ref{hrealeff}) 
could me modified by introducing a dependence on the spin variables.

In conclusion, the most remarkable feature of our improved EOB Hamiltonian 
is that in the extreme mass-ratio limit, it exactly reproduces the Hamiltonian of 
a spinning test particle in a Kerr spacetime, at linear order in the particle's 
spin and at all PN orders.

\begin{acknowledgments}
We thank Yi Pan for several useful discussions.
E.B. and A.B. acknowledge support from NSF Grants No. PHYS-0603762 and PHY-0903631.
  A.B. also acknowledges support from NASA grant NNX09AI81G. 
  \end{acknowledgments}

\appendix

\section{Incorporating spin-spin couplings in the effective-one-body Hamiltonian}
\label{sec:ss}

The Hamiltonian for a spinning particle in a Kerr spacetime 
that we derived in Sec.~\ref{sec:hamiltonian_kerr}, 
and the Hamiltonian for a spinning particle in a deformed-Kerr 
spacetime that we derived in Sec.~\ref{sec:kerrdeformed}
are only valid at linear order in the particle's spin.
However, as suggested in Ref.~\cite{Damour01c}, we can introduce
the terms that are quadratic in the particle's spin by modifying the 
quadrupole moment of the Kerr metric.

In particular, we can add a quadrupole which is quadratic in the particle's spin 
to the quadrupole of the Kerr metric (which is quadratic 
in $\boldsymbol{S}_{\rm Kerr}$).
The expression for the metric perturbation corresponding to a slight change of the Kerr quadrupole
can be extracted from the Hartle-Thorne metric~\cite{HTmetric1,HTmetric2}, which describes the spacetime 
of a slowly rotating star. Ref.~\cite{quasiKerr} gives this expression in quasi-Boyer-Lindquist coordinates
(i.e., in coordinates that reduce to Boyer-Lindquist coordinates if the quadrupole perturbation
is zero, thus reducing the spacetime to pure Kerr). This is exactly what is needed for our purposes,
since we work in quasi-Boyer-Lindquist coordinates too.

In particular, our procedure for introducing the terms which are
quadratic in the particle's spin into our Hamiltonian consists of
modifying the effective metric~\eqref{def_metric_in}--\eqref{def_metric_fin} by adding the
quadrupole metric
\begin{equation}
h^{\mu\nu} = \frac{1}{M^4} Q^{ij}\, S^\ast_i\, S^\ast_j\, \bar{h}^{\mu\nu}\,,
\end{equation}
%
where the quadrupole tensor $Q_{ij}$ is given by
\begin{equation}
Q_{ij}=\delta_{ij}-3 n_i\,n_j\,,
\end{equation}
and  $\bar{h}^{\mu\nu}$ is given by~\cite{quasiKerr}
\begin{eqnarray}
\bar{h}^{tt} &=& \frac{1}{1-{2M}/{r}}\, {\cal F}_1(r)\,, \quad \quad \bar{h}^{ti}=0\,,\\
\bar{h}^{ij}&=&-{\cal F}_{2}(r)\,\left\{\delta_{ij}-n_i\, n_j 
\left[1+\left(1-\frac{2M}{r}\right)\, \frac{{\cal F}_1(r)}{{\cal F}_{2}(r)}\right]\right\}\,.
\nonumber \\
\end{eqnarray}
The functions ${\cal F}_{1,2}(r)$ in the above equation are derived in Ref.~\cite{quasiKerr} by
transforming the Hartle-Thorne metric to quasi-Boyer-Lindquist coordinates, and are given by
\begin{eqnarray}
{\cal F}_{1}(r) &=& -\frac{5(r-M)}{8M\,r\,(r-2M)}\,(2M^2 + 6M\,r -3r^2) \nonumber \\
&& -\frac{15 r\,(r-2M)}{16 M^2}\,\log\left ( \frac{r}{r-2M} \right )\,,
\\
{\cal F}_{2}(r) &=& \frac{5}{8M\,r}\,(2M^2 -3M\,r -3r^2) + \nonumber \\
&& \frac{15}{16M^2} (r^2 -2M^2)\log \left ( \frac{r}{r-2M}\right ) \,.
\end{eqnarray}
Because at large radii ${\cal F}_{2}(r)\approx
-{\cal F}_{1}(r) \approx (M/r)^3$, we see that the deformation $h^{\mu\nu}$, when
inserted in the Hamiltonian~\eqref{eq:Hnsdef},
gives the correct leading-order (2PN) coupling of the particle's spin with itself. 
Keeping only the leading-order 
term created by $h^{\mu\nu}$, the effective EOB Hamiltonian therefore becomes
\begin{eqnarray}\label{full_H}
H &=& {H}_{\rm S} + \beta^i\,p_i + \alpha\,\sqrt{m^2 + \gamma^{ij}\,p_i\,p_j+{\cal Q}_4(p)} \nonumber \\
&& -\frac{m}{2M\, r^3}\,Q^{ij}\, S^\ast_i\, S^\ast_j\,.
\end{eqnarray}
\bibliography{references}

\end{document}